\documentclass[12pt,notoc]{JHEP3}

\usepackage{amsmath,amssymb,euscript,array,cite,mathrsfs,amsfonts}
\usepackage{epsfig}

\newcommand{\field}[1]{\mathbb{#1}} 
\newcommand*{\Dsl}[0]{{\rlap{\kern2.25pt /}{D}}}
\newcommand*{\Asl}[0]{{\rlap{\kern2.25pt /}{A}}}
\newcommand*{\dsl}[0]{{\rlap{\kern0.5pt /}{\partial}}}
\newcommand*{\xisl}[0]{{\rlap{\kern0.5pt /}{\xi}}}
\newcommand*{\asl}[0]{{\rlap{\kern0.5pt /}{a}}}
\newcommand*{\bsl}[0]{{\rlap{\kern0.5pt /}{b}}}

\newcommand{\lapprox}{\raisebox{-0.5ex}{$\
\stackrel{\textstyle<}{\textstyle\sim}\ $}}
\newcommand{\gapprox}{\raisebox{-0.5ex}{$\
\stackrel{\textstyle>}{\textstyle\sim}\ $}}

\def\Dslash{\,\,{\raise.15ex\hbox{/}\mkern-12mu D}}

\newcommand{\SP}[1]{\begin{equation}\begin{split} #1
\end{split}\end{equation}}

\newcommand{\Tr}{\operatorname{Tr}}

\def\B0{{\boldsymbol 0}}

\def\Tr{{\rm Tr}}

\def\det{{\rm det}}

\def\Dbarslash{\,\,{\raise.15ex\hbox{/}\mkern-12mu {\bar D}}}
\def\Dslash{\,\,{\raise.15ex\hbox{/}\mkern-12mu D}}
\def\delslash{\,\,{\raise.15ex\hbox{/}\mkern-9mu \partial}}
\def\delbarslash{\,\,{\raise.15ex\hbox{/}\mkern-9mu {\bar\partial}}}

\newcommand{\EQ}[1]{\begin{equation}\begin{split} #1
\end{split}\end{equation}}

\title{Bekenstein entropy bound for weakly-coupled field theories on a $3$-sphere}
\author{Joyce C. Myers\\ University of Groningen,
   Centre for Theoretical Physics, 9747 AG, Groningen, The Netherlands\\
\email{j.c.myers@rug.nl}}
\abstract{We calculate the high temperature partition functions for $SU(N_c)$ or $U(N_c)$ gauge theories in the deconfined phase on $S^1 \times S^3$, with scalars, vectors, and/or fermions in an arbitrary representation, at zero 't Hooft coupling and large $N_c$, using analytical methods. We compare these with numerical results which are also valid in the low temperature limit and show that the Bekenstein entropy bound resulting from the partition functions for theories with any amount of massless scalar, fermionic, and/or vector matter is always satisfied when the zero-point contribution is included, while the theory is sufficiently far from a phase transition. We further consider the effect of adding massive scalar or fermionic matter and show that the Bekenstein bound is satisfied when the Casimir energy is regularized under the constraint that it vanishes in the large mass limit. These calculations can be generalized straightforwardly for the case of a different number of spatial dimensions.}

\preprint{}

\begin{document}

\section{Introduction}

The possible existence of entropy bounds in weakly-coupled gravity theories which depend on the size of a compact space has been an ongoing subject of investigation since it was initiated by Bekenstein in \cite{Bekenstein:1980jp}. In this paper Bekenstein provided evidence for of a universal upper bound on the entropy over energy ratio for a weakly-coupled gravitational theory from plausibility arguments based on satisfying the generalized second law of thermodynamics. The bound takes the form
\EQ{
S \le 2 \pi R E \, ,
\label{bek_bnd}
}
where $S$ is the entropy, $E$ is the total energy, and $R$ is the effective radius of the system under consideration. One aspect which is particularly interesting about this inequality is that it is independent of Newton's constant $G$. In particular, the bound should hold in the limit $G \rightarrow 0$, and therefore it has also been tested in weakly-coupled field theories on $S^1 \times S^{d-1}$ in \cite{Kutasov:2000td,Klemm:2001db,Dowker:2002fd,Brevik:2002gh,Brevik:2002xq,Elizalde:2003cv,Gibbons:2006ij}, where $R$ can be simply interpreted as the radius of $S^{d-1}$. We extend these calculations, considering weakly-coupled field theories on $S^1 \times S^3$ with massive scalar and/or fermionic matter, using the technique in \cite{Cardy:1991kr,Kutasov:2000td} to obtain an analytical form of the partition functions in the high temperature limit, and compare with numerical results using the total partition function including the low temperature contributions. The calculations in \cite{Kutasov:2000td,Klemm:2001db,Dowker:2002fd,Brevik:2002gh,Brevik:2002xq,Elizalde:2003cv,Gibbons:2006ij} were performed for $U(1)$ theories, but generalization to $U(N_c)$ or $SU(N_c)$ theories, where $N_c$ is the number of colors, is straightforward: if the theory is in the deconfined phase such that interactions with the temporal gauge field can be neglected, then for fields in the representation ${\cal R}$ the partition function only differs by an overall factor of $d_{{\cal R}}$, which gives the dimension of the representation, and by the Jacobian contribution. We use the saddle point approximation to calculate the partition function from the action, which is valid in the large $N_c$ limit and gives a rough approximation for finite $N_c$.

Before presenting the calculations of the high temperature partition functions for massive scalar and fermionic matter, we provide a review of the calculations for massless scalars, fermions, and vectors to clarify how the Bekenstein bound is satisfied at all temperatures at zero 't Hooft coupling, while the theory is in the deconfined phase. This is accomplished by including the Casimir energy as stressed in \cite{Bekenstein:1980jp}. Our calculations provide an extension of \cite{Kutasov:2000td} to slightly lower temperatures. Regardless, we come to the same conclusion regarding ${\cal N} = 4$ Supersymmetric Yang-Mills theory and the Verlinde bound, if we consider the free theory, which is that it is eventually violated as the temperature is decreased from infinity, before the deconfinement-confinement transition can take place. However, for the $SU(N_c)$ theory there is a contribution from the Jacobian factor which could prevent violation of the Verlinde bound. Our small extension to lower temperatures is obtained by retaining an additional constant term in the high temperature expansion of the massless scalar partition function which corresponds to a constant of integration in the calculation of \cite{Kutasov:2000td} and which can be neglected at sufficiently high temperatures \footnote{It should be noted that this constant should not be dropped to match onto the low temperature result where the requirement from the third law of thermodynamics is that the entropy is zero at zero temperature. The constant is relevant at intermediate temperatures, and the expansion is no longer valid at low temperatures.}. In addition, the inclusion of the Casimir energies is significant in satisfying the Bekenstein bound at lower temperatures. These considerations, along with numerical calculations to check the partition functions at low temperatures, bring us to slightly modify the conclusion of \cite{Kutasov:2000td} regarding the Bekenstein bound, in that we find it to be satisfied at all temperatures for all free theories with any number of massless scalars, fermions, and/or vectors, in four dimensions. Therefore, we also slightly modify the conclusion of \cite{Klemm:2001db,Brevik:2002gh,Brevik:2002xq}, where the absence of the constant term in the partition function led to the conclusion that the entropy over energy ratio diverges to positive infinity for ${\cal N} = 4$ SYM theory at low temperature.

Following the calculations of the high temperature partition functions for theories with massless matter we consider the analogous calculations for theories with massive scalar and fermionic matter. Contrary to the case with massless fields where the high temperature expansion is obtained by considering a closed contour integral over a finite number of simple poles, the calculation with massive matter receives an infinite number of residues and the high temperature partition function contains an infinite sum which is convergent for $m \beta \le 2 \pi$ in the case of scalars, where $m$ is the mass and $\beta$ is the inverse temperature, and for $m \beta \le \pi$ in the case of fermions. For masses above these values the sums would need to be regularized. Rather than carrying out the regularizations to extrapolate the results to large $m \beta$ we present numerical calculations using the original full partition function, which are valid for any $m \beta$. It is worth mentioning that calculations relevant to testing the Bekenstein bound for theories with massive scalar matter were also considered in \cite{Elizalde:2003cv} where the high temperature partition function is obtained through explicit evaluation of the integrals in the Abel-Plana formula in place of the Mellin Transform and contour integral approach. The partition functions calculated using these techniques agree in the small $m R$ limit, up to $\beta$-independent constants, which are important in extending the results to slightly lower temperatures. 

The layout of the paper is as follows. Section \ref{bek_bnd_calc} describes how we will test the Bekenstein bound. Section \ref{logZreview} shows how the partition functions are obtained from $1$-loop perturbation theory. Section \ref{cas_rev} shows the calculations for the Casimir energies. Section \ref{highTlogZ} gives the calculations of the high temperature partition functions for fields in the deconfined phase and shows the calculations that test the Bekenstein bound. Section \ref{lowTconf_sec} discusses what happens if the theory is in the confined phase. Section \ref{Verlinde_sec} shows the calculations to test the Verlinde bound for ${\cal N} = 4$ SYM theory. And, section \ref{conclusions_sec} reports the conclusions.

\subsection{Bekenstein bounds}
\label{bek_bnd_calc}

For theories formulated on $S^1 \times S^{d-1}$ the Bekenstein bound relates the maximum possible entropy to the total energy according to the relationship in (\ref{bek_bnd}) such that
\EQ{
L \equiv 2 \pi R E - S \ge 0 \, ,
}
where $S$ is the entropy given by
\EQ{
S = \beta (E - F) \, ,
}
in terms of the inverse temperature $\beta = \frac{1}{T}$, the total energy
\EQ{
E = - \frac{\partial}{\partial \beta} \log Z \, ,
}
where $Z$ is the partition function, and the free energy
\EQ{
F = -T \log Z \, .
}
Therefore the Bekenstein bound takes the equivalent form
\EQ{
L \equiv (\beta - 2 \pi R) \frac{\partial}{\partial \beta} \log Z - \log Z \ge 0 \, ,
\label{Lbek}
}
which shows that $L$ is an additive quantity, at zero 't Hooft coupling, such that if $L \ge 0$ is true separately for scalars, fermions, and vectors, then it is true for all theories which contain a combination of these fields.

Following \cite{Gibbons:2006ij} we will test the validity of the Bekenstein bound by obtaining the minimum of $L$ when possible. The derivatives of $L$ with respect to $\beta$,
\SP{
\frac{\partial L}{\partial \beta} &= (\beta - 2 \pi R) \frac{\partial^2}{\partial \beta^2} \log Z \, , \\
\frac{\partial^2 L}{\partial \beta^2} &= \frac{\partial^2}{\partial \beta^2} \log Z + (\beta - 2 \pi R) \frac{\partial^3}{\partial \beta^3} \log Z \, ,
}
indicate that $L$ has a minimum at $\beta = 2 \pi R$ when $\frac{\partial^2}{\partial \beta^2} \log Z > 0$. In this case satisfaction of the Bekenstein bound implies that
\EQ{
-\log Z \Big{|}_{\beta = 2 \pi R} \ge 0 \, ,
}
or equivalently that
\EQ{
F \Big{|}_{\beta = 2 \pi R} \ge 0 \, .
}
Following \cite{Cardy:1991kr,Kutasov:2000td,Gibbons:2006ij} we derive the high temperature partition functions for theories with massless scalars, vectors, and fermions, then consider theories with massive fermions and scalars. The calculations are carried out for $d=3$ spatial dimensions but they proceed in a similar manner for any $d$. In each case we evaluate $L$ using (\ref{Lbek}) to determine if the Bekenstein bound is satisfied. It turns out that this depends on whether the Casimir (zero-point) energy contribution is included in $\log Z$, and, in the case of massive matter, on how it is regularized.
\section{Partition functions}
\label{logZreview}

Although it is possible to obtain the partition functions at zero 't Hooft coupling from counting arguments we summarize how they are calculated from one-loop perturbation theory, following \cite{Aharony:2003sx}, as a physical motivation, and as a basis for considering interacting theories. The one-loop results are valid at all temperatures when the radius of the $S^3$ is much less than the inverse strong coupling scale, $R << \Lambda_{YM}^{-1}$. The contribution to the partition function from $N_s$ real scalars with mass $m$ is \footnote{Here we use the saddle point approximation to relate the action to the partition function via $S = - \log Z$ such that the integrals over $A_{\mu}$ evaluate to the stationary point solution. This is valid, for example, for $SU(N_c)$ or $U(N_c)$ theories at large $N_c$.}
\EQ{
\log Z_s = -\frac{N_s}{2} \log \det \left[ - D_0^2 - \Delta^{(s)} + m^2 \right] \, ,
\label{logZs1}
}
where $\Delta^{(s)}$ is the (conformally coupled) scalar Laplacian. Its eigenvalues $\varepsilon^{(s)}_l$ and degeneracies $d_l^{(s)}$ on $S^3$ with radius $R$ take the form
\SP{
\Delta^{(s)} Y_l ({\hat \Omega}) &= - \varepsilon_l^{(s) 2} Y_l ({\hat \Omega}) \, ,\\
\varepsilon_l^{(s)} &= \tfrac{1}{R} (l+1) \, ,\\
d_l^{(s)} &= (l+1)^2 \, ,
\label{eigens}
}
where $l = 0, 1, 2, ...$.

For vector theories, the spatial gauge field can be decomposed as $A_i = B_i + C_i$ where $B_i$ give the transverse components with $\nabla_i B_i = 0$, and $C_i$ give the longitudinal components with $C_i = \nabla_i f$. The temporal gauge field is decomposed as $A_0 = \alpha + g {\cal A}_0$, where the background field $\alpha$ contains the diagonal elements of $A_0$ and $g {\cal A}_0$ contains the off-diagonal elements. Then, keeping the one-loop contributions, the $C_i$ cancel almost completely against the ghost contribution and the off-diagonal fluctuations of the temporal gauge field. What remains is the Vandermonde contribution
\EQ{
\log Z_{Vdm} = \frac{1}{2} \log \det_{l=0} \left( - D_0^2 - \Delta^{(s,{\rm min})} \right) \, ,
}
which is the Jacobian factor necessary for converting between a unitary matrix and its eigenvalue angles. The relevant energy eigenvalues $\varepsilon^{(s,{\rm min})}_l$ and degeneracies $d_l^{(s)}$ of the (minimally coupled) scalar Laplacian $\Delta^{(s,{\rm min})}$ on $S^3$ are
\SP{
\Delta^{(s,{\rm min})} Y_l ({\hat \Omega}) &= - \varepsilon_l^{(s,{\rm min}) 2} Y_l ({\hat \Omega}) \, ,\\
\varepsilon_l^{(s,{\rm min})2} &= \tfrac{1}{R} l (l+2) \, ,\\
d_l^{(s)} &= (l+1)^2 \, ,
}
where $l = 0, 1, 2, ...$.

The contribution of the remaining (transverse) vectors $B^i$, to the partition function is
\EQ{
\log Z_v = - \frac{1}{2} \log \det \left[ - D_0^2 - \Delta^{(v)} \right] \, .
\label{logZv1}
}
The eigenvalues $\varepsilon^{(v)}_l$ and degeneracies $d_l^{(v)}$ of the transverse vector Laplacian $\Delta^{(v)}$ on $S^3$ are
\SP{
\Delta^{(v)} B^i_l ({\hat \Omega}) &= - \varepsilon_l^{(v) 2} B^i_l ({\hat \Omega}) \, ,\\
\varepsilon_l^{(v)} &= \tfrac{1}{R} (l+1) \, ,\\
d_l^{(v)} &= 2 l (l+2) \, ,
\label{eigenv}
}
where $l = 1, 2, ...$.

The contribution of $N_f$ (Majorana) fermions of mass $m$ is
\EQ{
\log Z_f = N_f \log \det \left[ - D_0^2 - \Delta^{(f)} + \frac{1}{4} {\mathscr R} + m^2 \right] \, ,
\label{logZf1}
}
where ${\mathscr R}$ is the scalar curvature of $S^3$. The eigenvalues $\varepsilon^{(f)}_l$ and degeneracies $d_l^{(f)}$ of the fermion Laplacian $\Delta^{(f)}$ on $S^3$ are
\SP{
\left( \Delta^{(f)} - \frac{1}{4} {\mathscr R} \right) \psi &= - \varepsilon_l^{(f) 2} \psi \, ,\\
\varepsilon_l^{(f)} &= \tfrac{1}{R} (l+\tfrac{1}{2}) \, ,\\
d_l^{(f)} &= l (l+1) \, ,
\label{eigenf}
}
where $l = 1, 2, ...$.

In terms of the constant background temporal gauge field, $A_{\mu} = \delta_{\mu 0} \alpha$, the partition functions for fields in the representation ${\cal R}$ can be calculated by using the decomposition $D_0 = \partial_0 + i \alpha \rightarrow i \omega_n^{\pm} + i \alpha$, then taking the determinants over momentum space to obtain
\SP{
\det (-D_0^2 + \varepsilon^2)_{\pm} &= \det \left( - (\partial_0 + i \alpha)^2 + \varepsilon^2 \right) \, ,\\
&= \det_{{\cal R}} \prod_{l} \prod_{n \in {\field Z}} \left( (\omega_n^{\pm} + \alpha)^2 + \varepsilon_l^2 + m^2 \right)^{d_l} \, ,
}
where the Matsubara frequencies are $\omega_n^+ = \frac{2 \pi n}{\beta}$ for fields with periodic boundary conditions around $S^1$, and $\omega_n^{-} = \frac{(2 n +1)\pi}{\beta}$ for fields with antiperiodic boundary conditions. Following \cite{Aharony:2003sx} for fields with periodic boundary conditions
\SP{
&\prod_{n \in {\field Z}} \left( (\omega_n^{+} + \alpha)^2 + \varepsilon^2 \right)\\
&= (\alpha^2 + \varepsilon^2) \prod_{k \ne 0} \left[ \frac{4 \pi^2 k^2}{\beta^2} \right] \prod_{n = 1}^{\infty} \left[ \left( 1 - \frac{\beta^2(\alpha + i \varepsilon)^2}{4 \pi^2 n^2} \right) \left( 1 - \frac{\beta^2(\alpha - i \varepsilon)^2}{4 \pi^2 n^2} \right) \right] \, ,\\
&= \frac{4}{\beta^2} \prod_{k = 1}^{\infty} \left[ \frac{16 \pi^4 k^4}{\beta^4} \right] \sin \left[ \frac{\beta}{2} (\alpha + i \varepsilon) \right] \sin\left[ \frac{\beta}{2} (\alpha - i \varepsilon) \right] \, ,\\
&= {\cal N} e^{\beta \varepsilon} \left( 1 - e^{-\beta \varepsilon + i \beta \alpha} \right) \left( 1 - e^{-\beta \varepsilon - i \beta \alpha} \right) \, ,
}
where the identity $\prod_{n=1}^{\infty} \left( 1 - \frac{x^2}{n^2} \right) = \frac{\sin (\pi x)}{\pi x}$ was used to obtain the third line, and
\EQ{
{\cal N} \equiv \frac{1}{\beta^2} \prod_{k = 1}^{\infty} \left[ \frac{16 \pi^4 k^4}{\beta^4} \right] \, .
}
Following the same procedure for fields with antiperiodic boundary conditions around $S^1$ results in the replacement $\alpha \rightarrow \alpha + \frac{\pi}{\beta}$. That is,
\EQ{
\prod_{n \in {\field Z}} \left( (\omega_n^{-} + \alpha)^2 + \varepsilon^2 \right) = {\cal N} e^{\beta \varepsilon} \left( 1 + e^{-\beta \varepsilon + i \beta \alpha} \right) \left( 1 + e^{-\beta \varepsilon - i \beta \alpha} \right) \, .
}
Putting it all together, shifting $\alpha$ to the new independent variable $\theta \equiv \beta \alpha$ such that $\theta = {\rm diag} \{ \theta_1, \theta_2, ..., \theta_{N_c} \}$ is the matrix of eigenvalue angles of the Polyakov lines ${\mathscr P}_n^{({\cal R})} \equiv \Tr_{{\cal R}} e^{i n \beta \alpha} = \Tr_{{\cal R}} e^{i n \theta}$, gives the result
\EQ{
\log \det (-D_0^2 + \varepsilon^2)_{\pm} = d_{{\cal R}} \beta \sum_{l} d_l \varepsilon_l - \sum_{n=1}^{\infty} \frac{(\pm 1)^n}{n} \sum_l d_l e^{-n \beta \varepsilon_l} \left[ \Tr_{{\cal R}} e^{i n \theta} + \Tr_{{\cal R}} e^{-i n \theta} \right] \, ,
\label{logdet}
}
for fields in the representation ${\cal R}$, with dimension $d_{{\cal R}}$. The top sign in (\ref{logdet}) is for fields with periodic boundary conditions around $S^1$ and the bottom sign is for fields with antiperiodic boundary conditions. The absence of an ${\cal N}$-dependent term in (\ref{logdet}) results from
\SP{
\log {\cal N} &= -2 \log \beta + 4 \sum_{k=1}^{\infty} \left[ \log \left(\frac{2 \pi}{\beta}\right) + \log k \right] \, ,\\
&= -2 \log \beta + 4 \log \left(\frac{2 \pi}{\beta}\right) \zeta(0) - 4 \zeta'(0) \, ,\\
&= 0 \, .
}
Using (\ref{logdet}) the Vandermonde contribution becomes
\SP{
\log Z_{Vdm} &= \frac{1}{2} d_{{\cal A}} \beta d_0 \varepsilon_0 - \sum_{n=1}^{\infty} \frac{1}{n} d_0 e^{-n \beta \varepsilon_0} \Tr_{{\cal A}} e^{i n \theta} \, ,\\
&= - \sum_{n=1}^{\infty} \frac{1}{n} \Tr_{{\cal A}} e^{i n \theta} \, ,
\label{vdm}
}
where $d_{{\cal A}}$ is the dimension of the adjoint representation. In the calculations of the high temperature partition functions that follow we drop this contribution. From (\ref{vdm}) one finds that in the high temperature deconfined phase $\Tr_{{\cal A}} e^{i n \theta} \rightarrow d_{{\cal A}}$ and $\log Z_{Vdm} \rightarrow - d_{{\cal A}} \zeta(1)$. In the low temperature confined phase, $\Tr_{{\cal A}} e^{i n \theta} \rightarrow 0$ for $U(N_c)$ vectors, and $\Tr_{{\cal A}} e^{i n \theta} \rightarrow -1$ for $SU(N_c)$ vectors such that $\log Z_{Vdm} \rightarrow 0$ or $\zeta(1)$, respectively. For the $SU(N_c)$ theory it is necessary to shift the temperature behavior of $\log Z_{Vdm}$ from $\zeta(1)$ to zero such that the entropy $S \rightarrow 0$ at zero temperature, following \cite{Bekenstein:1980jp}, to satisfy the third law of thermodynamics. The high temperature contribution should be considered more carefully. In order that the free energy doesn't slowly diverge we impose the requirement that $\lim_{n \rightarrow \infty} \Tr_{{\cal A}} e^{i n \theta} = 0$. This allows the sum to be truncated. In the high temperature limit the result is $\log Z_{Vdm} \simeq - d_{{\cal A}} \sum_{n = 1}^{\Lambda} \frac{1}{n} \simeq - d_{{\cal A}} ( \log \Lambda + \gamma_E )$, where $\Lambda$ is large. This truncated result can be dominated by contributions from the single particle partition functions for the matter fields in the high temperature limit, allowing for the recovery of the expected one-loop free energy and other thermodynamic observables as $R \rightarrow \infty$. Using (\ref{Lbek}) the Vandermonde term contributes favorably towards satisfaction of the Bekenstein bound with $L_{Vdm} = - \log Z_{Vdm} \simeq d_{{\cal A}} ( \log \Lambda + \gamma_E )$. In what follows we will ignore this contribution since $\Lambda$ has an unknown temperature dependence, but we keep in mind that it can only help the Bekenstein bound to be satisfied. Dropping this term entirely is valid deep in the deconfined phase in the very high temperature limit (The Vandermonde term can be included in the calculation of the vector partition function as in \cite{Aharony:2003sx}. The partition function can then be precisely calculated at very high temperatures but numerical checks are not possible.). If we wished to study the theory close to a deconfinement-confinement transition then consideration of the $\theta$-dependence of the Vandermonde piece would be crucial.
\section{Casimir Energy}
\label{cas_rev}

The Casimir energies for real scalars and Weyl fermions in the representation ${\cal R}$, and adjoint vectors, are given by, respectively,
\SP{
E_{Cas}^{(s)} &= \frac{d_{{\cal R}}^{(s)}}{2 R} \sum_l d_l^{(s)} \varepsilon_l^{(s)} \, , \\
E_{Cas}^{(f)} &= - \frac{d_{{\cal R}}^{(f)}}{R} \sum_l d_l^{(f)} \varepsilon_l^{(f)} \, , \\
E_{Cas}^{(v)} &= \frac{d_{{\cal A}}}{2 R} \sum_l d_l^{(v)} \varepsilon_l^{(v)} \, .
}
To obtain the Casimir energies it is necessary to regulate the sums. When the matter is massless it is straightforward to perform these sums using zeta function regularization. When it is massive, it is unclear how to obtain the appropriately regularized result. One option is to obtain results using a cutoff regularization scheme and compare with those using zeta function regularization to define an undetermined regularization parameter and obtain a, perhaps scheme independent, result. Another possibility is to use only zeta function regularization then apply a physical constraint to obtain the undetermined regularization parameter, namely, considering the masses as quantum corrections such that their contribution to the energy is constrained to vanish as the mass is taken to infinity.
\subsection{Scalars}

For real scalars with mass $m$ the Casimir energy is
\SP{
E_{Cas}^{(s)} &= \frac{d_{{\cal R}}^{(s)}}{2 R} \sum_{l=0}^{\infty} (l+1)^2 \sqrt{(l+1)^2 + m^2 R^2} \, .
}
In the massless limit it is straightforward to obtain the Casimir energy using zeta function regularization,
\SP{
E_{Cas}^{(s)} \Big|_{m R = 0} &= \frac{d_{{\cal R}}^{(s)}}{2 R} \sum_{l=1}^{\infty} l^3 = \frac{d_{{\cal R}}^{(s)}}{2 R} \zeta(-3) = \frac{d_{{\cal R}}^{(s)}}{240 R} \, .
\label{scacas0}
}
When $m R \ne 0$ it is more complicated to regularize the sum. First we follow \cite{Lim:2008yv} and consider two regularization schemes: cutoff regularization, and zeta function regularization. We match the zeta function regularized result against the cutoff-independent part of the cutoff-regularized result to obtain the appropriate normalization. Second, we calculate the Casimir energy using the constraint that it vanishes as the mass goes to infinity reproducing the results in \cite{Elizalde:2003cv,Elizalde:2003wd}. In both cases the zeta function regularization procedure in \cite{Blau:1988kv} is used.
\subsubsection{Cutoff regularization}

Following \cite{Lim:2008yv} we define the cutoff regularized Casimir energy for a real scalar field as
\EQ{
E_{Cas}^{(s)}(\lambda) \equiv - \frac{d_{{\cal R}}^{(s)}}{2 R} \frac{\partial}{\partial \lambda} \sum_{l=0}^{\infty} (l+1)^2 e^{-\lambda \sqrt{(l+1)^2 + m^2 R^2}} \, .
\label{cutoff-cas-sca}
}
To solve the sum it is useful to express the exponential as a contour integral via the Mellin Transform
\EQ{
e^{-x} = \frac{1}{2 \pi i} \int_{{\cal C}} {\rm d}s \, \Gamma(s) x^{-s} \, ,
\label{mellin}
}
where the contour ${\cal C}$ extends from $c_0 - i \infty$ to $c_0 + i \infty$ with $c_0 >  3$ (for $d$ dimensions $c_0 > d-1$) to allow the sum over $l$ to be brought into the integral. The Casimir energy is then
\EQ{
E_{Cas}^{(s)}(\lambda) = - \frac{d_{{\cal R}}^{(s)}}{2 R} \frac{\partial}{\partial \lambda} \frac{1}{2 \pi i} \int_{{\cal C}} {\rm d}s \, G_s (s) \, ,
\label{sca_cas_contour}
}
with
\SP{
G_s (s) &= \lambda^{-s} \Gamma(s) \sum_{l=0}^{\infty} (l+1)^2 [(l+1)^2 + m^2 R^2]^{-s/2}\\
&= \lambda^{-s} \Gamma(s) \sum_{l=1}^{\infty} \bigg[ \left[l^2 + m^2 R^2 \right]^{1-s/2} - m^2 R^2 \left[l^2 + m^2 R^2 \right]^{-s/2} \bigg] \, .
}
The sum over $l$ can be solved by zeta function regularization using (1.38) in \cite{Elizalde:1996zk} (see also \cite{Elizalde:2003cv,Elizalde:2003wd}),
\SP{
\sum_{l=1}^{\infty} \left( l^2 + M^2 \right)^{-s} = &- \frac{M^{-2 s}}{2} + \frac{\sqrt{\pi}}{2} M^{-2 s + 1} \frac{\Gamma(s - \frac{1}{2})}{\Gamma(s)}\\
&+ \frac{2 \pi^s}{\Gamma(s)} M^{-s+1/2} \sum_{n=1}^{\infty} n^{s-1/2} K_{s-1/2} (2 \pi n M) \, .
\label{zeta_scalar}
}
Note that the sum converges to this result for $s > \frac{1}{2}$ and this is why the contour was required to have $c_0 > 3$. (\ref{zeta_scalar}) also gives the appropriate analytic continuation for other $s$. Using this result gives
\SP{
G_s (s) = & \lambda^{-s} 2^{s} \Gamma(\tfrac{s+1}{2}) \bigg[ \frac{1}{8} (m R)^{3-s} \Gamma(\tfrac{s-3}{2}) + \sum_{n=1}^{\infty} \bigg[ (\tfrac{s}{2} - 1) \left( \frac{\pi n}{m R} \right)^{(s-3)/2} K_{\frac{s-3}{2}} (2 \pi n m R)\\
&\hspace{24mm} - m^2 R^2 \left( \frac{\pi n}{m R} \right)^{(s-1)/2} K_{\frac{s-1}{2}} (2 \pi n m R) \bigg] \bigg] \, .
}
The next step is to add an arc at infinity in the left-hand complex $s$-plane to close the contour in (\ref{sca_cas_contour}) around the poles in $G_s(s)$ and collect the residues. The contribution of this arc alone to the contour integral is negligible for sufficiently small values of $\lambda$. Evaluating $G_s(s)$ at the $s$-values corresponding to singularities in the Gamma functions gives rise to the relevant simple poles
\SP{
&G_s(s \rightarrow 3) \propto \left( \frac{1}{s-3} \right) \frac{2}{\lambda^3} \, ,\\
&G_s(s \rightarrow 1) \propto - \left( \frac{1}{s-1} \right) \frac{m^2 R^2}{2 \lambda} \, ,\\
&G_s(s \rightarrow -1) \propto \left( \frac{1}{s+1} \right) \lambda \, \Bigg{\{} m^4 R^4 \bigg[ \frac{3}{32} - \frac{1}{8} \gamma_E + \frac{1}{8} \log \left( \frac{2}{\lambda m R} \right) \bigg]\\
&\hspace{24mm} - \sum_{n=1}^{\infty} \left[ \frac{3}{2} \left( \frac{m R}{\pi n} \right)^2 K_{2} (2 \pi n m R) + m^2 R^2 \left( \frac{m R}{\pi n} \right) K_{1} (2 \pi n m R) \right] \Bigg{\}} \, .
}
The remaining simple poles (for $s = -3, -5, ...$) are higher order in $\lambda$ and produce negligible, cutoff-dependent contributions, even after taking the $\lambda$-derivative in (\ref{cutoff-cas-sca}). Collecting the residues the cutoff-regularized Casimir energy takes the final form
\SP{
E_{Cas}^{(s)}(\lambda) \simeq &\frac{d_{{\cal R}}^{(s)}}{2 R} \Bigg{\{} \frac{6}{\lambda^4} - \frac{m^2 R^2}{2 \lambda^2} + m^4 R^4 \bigg[ \frac{1}{32} + \frac{1}{8} \gamma_E + \frac{1}{8} \log \left( \frac{\lambda m R}{2} \right) \bigg]\\
&+ \sum_{n=1}^{\infty} \left[ \frac{3}{2} \left( \frac{m R}{\pi n} \right)^2 K_{2} (2 \pi n m R) + m^2 R^2 \left( \frac{m R}{\pi n} \right) K_{1} (2 \pi n m R) \right] \Bigg{\}} \, .
\label{cutoff-cas-final-sca}
}
\subsubsection{Zeta function regularization}

To obtain the zeta function regularized Casimir energy we follow the scheme in \cite{Blau:1988kv}. The Casimir energy is defined as
\EQ{
E_{Cas}^{(s)} (\mu) = \frac{d_{{\cal R}}^{(s)}}{2 R} \frac{1}{2} \lim_{\varepsilon \rightarrow 0} \left[ \zeta_p (-\tfrac{1}{2} + \varepsilon) + \zeta_p (- \tfrac{1}{2} - \varepsilon) \right] \, ,
}
where
\EQ{
\zeta_p (s) \equiv \mu^{1+2s} \sum_{l=0}^{\infty} (l+1)^2 \left[ (l+1)^2 + m^2 R^2 \right]^{-s} \, ,
}
and $\mu$ is a normalization factor to be determined later. Dividing the sum into parts which can be separately regularized using zeta function techniques gives
\EQ{
\zeta_p(s) = \mu^{1+2s} \sum_{l=1}^{\infty} \bigg[ \left[l^2 + m^2 R^2 \right]^{1-s} - m^2 R^2 \left[l^2 + m^2 R^2 \right]^{-s} \bigg] \, .
}
Then, using (\ref{zeta_scalar}), the Casimir energy takes the $\mu$-dependent form
\SP{
E_{Cas}^{(s)} (\mu) = &\frac{d_{{\cal R}}^{(s)}}{2 R} \Bigg{\{} m^4 R^4 \left[ \frac{1}{32} + \frac{1}{8} \log \left( \frac{m R}{2 \mu} \right) \right]\\
&+ \sum_{n=1}^{\infty} \left[ \frac{3}{2} \left( \frac{m R}{\pi n} \right)^2 K_{2} (2 \pi n m R) + m^2 R^2 \left( \frac{m R}{\pi n} \right) K_{1} (2 \pi n m R) \right] \Bigg{\}} \, .
}
Comparing this result with the cutoff-independent contributions in (\ref{cutoff-cas-final-sca}) we obtain the value of the finite normalization factor, $\mu \equiv e^{- \gamma_E}$, as found in the system in \cite{Lim:2008yv}. We refer to the $\mu \equiv e^{- \gamma_E}$-regularized Casimir energy as the scheme $I$ result, given by
\SP{
E_{Cas}^{(s)I} = &\frac{d_{{\cal R}}^{(s)}}{2 R} \Bigg{\{} m^4 R^4 \left[ \frac{1}{32} + \frac{1}{8} \gamma_E + \frac{1}{8} \log \left( \frac{m R}{2} \right) \right]\\
&+ \sum_{n=1}^{\infty} \left[ \frac{3}{2} \left( \frac{m R}{\pi n} \right)^2 K_{2} (2 \pi n m R) + m^2 R^2 \left( \frac{m R}{\pi n} \right) K_{1} (2 \pi n m R) \right] \Bigg{\}} \, .
\label{scacas1}
}
Another way to regularize the Casimir energy is to impose the physically-motivated constraint that the matter contribution should vanish in the large mass limit \cite{Bordag:2001qi}. The terms which diverge as $m R \rightarrow \infty$ can be removed by an appropriate definition of $\mu$. In this case $\mu \equiv \frac{1}{2} e^{1/4} m R$, and the resulting Casimir energy is referred to as the scheme $II$ result, given by
\SP{
E_{Cas}^{(s)II} = &\frac{d_{{\cal R}}^{(s)}}{2 R} \sum_{n=1}^{\infty} \left[ \frac{3}{2} \left( \frac{m R}{\pi n} \right)^2 K_{2} (2 \pi n m R) + m^2 R^2 \left( \frac{m R}{\pi n} \right) K_{1} (2 \pi n m R) \right] \, ,
\label{scacas2}
}
in agreement with \cite{Elizalde:2003cv,Elizalde:2003wd}.

A few comments are in order with regard to choosing a regularization scheme. It is clear that a scalar theory regularized according to scheme $I$ results in $E_{Cas} < 0$ for a range of $m R$. As discussed in \cite{Bekenstein:1980jp} this leads to a partition function that can violate the entropy bound (\ref{bek_bnd}). This is clear by considering the large $\frac{\beta}{R}$ limit where the only remaining contribution to the partition function is that of the Casimir energy. Then $L = -\frac{2 \pi R}{\beta} \log Z = 2 \pi R E_{Cas}$, which violates the Bekenstein bound when $E_{Cas} < 0$. However, if scheme $II$ is chosen then $E_{Cas} > 0$ for all $m R$ and the Bekenstein bound is satisfied in this limit.
\subsection{Fermions}
For Weyl (Majorana) fermions with mass $m$ the Casimir energy is
\EQ{
E_{Cas}^{(f)} = - \frac{d_{{\cal R}}^{(f)}}{R} \sum_{l=0}^{\infty} l (l+1) \sqrt{(l+\tfrac{1}{2})^2 + m^2 R^2} \, .
}
In the massless limit the Casimir energy is obtained by ordinary zeta function regularization, such that
\EQ{
E_{Cas}^{(f)} \Big|_{m R = 0} = - \frac{d_{{\cal R}}^{(f)}}{R} \sum_{l=0}^{\infty} l (l+1) (l+\tfrac{1}{2}) = - \frac{d_{{\cal R}}^{(f)}}{R} \left[ \zeta(-3,\tfrac{1}{2}) - \frac{1}{4} \zeta(-1,\tfrac{1}{2}) \right] = \frac{17 d_{{\cal R}}^{(f)}}{960 R} \, .
\label{fercas0}
}
For $m R \ne 0$ we proceed as for scalars and first obtain a regularized Casimir energy by comparing results from cutoff regularization and zeta function regularization, and second by using only zeta function regularization and imposing the constraint\\ $\lim_{m R \rightarrow \infty} E_{Cas} = 0$.
\subsubsection{Cutoff regularization}

Following the procedure for scalars the cutoff regularized Casimir energy for fermions is
\SP{
E_{Cas}^{(f)}(\lambda) &\equiv \frac{d_{{\cal R}}^{(f)}}{R} \frac{\partial}{\partial \lambda} \sum_{l=0}^{\infty} l (l+1) e^{-\lambda \sqrt{(l+\frac{1}{2})^2 + m^2 R^2}} \, , \\
&= \frac{d_{{\cal R}}^{(f)}}{R} \frac{\partial}{\partial \lambda} \frac{1}{2 \pi i} \int_{{\cal C}} {\rm d}s \, G_f (s) \, ,
\label{cutoff-cas-fer}
}
with
\SP{
G_f (s) &= \lambda^{-s} \Gamma(s) \sum_{l=0}^{\infty} l (l+1) [(l+\tfrac{1}{2})^2 + m^2 R^2]^{-s/2} \, ,\\
&= \lambda^{-s} \Gamma(s) \sum_{l=0}^{\infty} \bigg[ \left[(l+\tfrac{1}{2})^2 + m^2 R^2 \right]^{1-s/2} - (\tfrac{1}{4} + m^2 R^2) \left[(l+\tfrac{1}{2})^2 + m^2 R^2 \right]^{-s/2} \bigg] \, .
}
The sum over $l$ can be solved by zeta function regularization using (4.17) in \cite{Elizalde:1996zk},
\SP{
&\sum_{l=0}^{\infty} \left[ (l + \tfrac{1}{2})^2 + M^2 \right]^{-s}\\
&= \frac{\sqrt{\pi}}{2} \frac{\Gamma(s-\tfrac{1}{2})}{\Gamma(s)} |M|^{1-2s} + \frac{2 \pi^s}{\Gamma(s)} |M|^{1/2-s} \sum_{n=1}^{\infty} (-1)^n n^{s-1/2} K_{s-1/2} (2 \pi n |M|) \, ,
\label{zeta_fermion}
}
where the sum converges for all $s > \frac{1}{2}$ and the result provides the appropriate analytic continuation for other $s$. Using this result gives
\SP{
G_f (s) = & \lambda^{-s} 2^{s} \Gamma(\tfrac{s+1}{2}) \bigg[ \frac{1}{8} (m R)^{1-s} \left( m^2 R^2 - \frac{1}{4} (s - 3) \right) \Gamma(\tfrac{s-3}{2})\\
&+ \sum_{n=1}^{\infty} (-1)^n \bigg[ (\tfrac{s}{2} - 1) \left( \frac{\pi n}{m R} \right)^{(s-3)/2} K_{\frac{s-3}{2}} (2 \pi n m R)\\
&\hspace{24mm} - (\tfrac{1}{4} + m^2 R^2) \left( \frac{\pi n}{m R} \right)^{(s-1)/2} K_{\frac{s-1}{2}} (2 \pi n m R) \bigg] \bigg] \, .
}
Evaluating $G_f(s)$ at the $s$-values corresponding to singularities in the Gamma functions gives rise to the relevant simple poles at $s = 3, 1, -1$. The remaining simple poles (for $s = -3, -5, ...$) are higher order in $\lambda$ and produce negligible, cutoff-dependent contributions. Collecting the residues the cutoff-regularized Casimir energy takes the final form
\SP{
E_{Cas}^{(f)}(\lambda) \simeq &\frac{d_{{\cal R}}^{(f)}}{R} \Bigg{\{} \frac{6}{\lambda^4} - \frac{(m^2 R^2 + \frac{1}{2})}{2 \lambda^2} + m^2 R^2 \bigg[ \frac{1}{16} + \frac{1}{32} m^2 R^2  + \frac{1}{8} (1 + m^2 R^2) \gamma_E\\
&+ \frac{1}{8} (1 + m^2 R^2) \log \left( \frac{\lambda m R}{2} \right) \bigg]\\
&+ \sum_{n=1}^{\infty} (-1)^n \left[ \frac{3}{2} \left( \frac{m R}{\pi n} \right)^2 K_{2} (2 \pi n m R) + (m^2 R^2 + \tfrac{1}{4}) \left( \frac{m R}{\pi n} \right) K_{1} (2 \pi n m R) \right] \Bigg{\}} \, .
\label{cutoff-cas-final-fer}
}
\subsubsection{Zeta function regularization}

To obtain the zeta function regularized Casimir energy we again follow the regularization prescription in \cite{Blau:1988kv}, where for fermions
\EQ{
E_{Cas}^{(f)}(\mu) = - \frac{d_{{\cal R}}^{(f)}}{2 R} \lim_{\varepsilon \rightarrow 0} \left[ \zeta_p (-\tfrac{1}{2} + \varepsilon) + \zeta_p (- \tfrac{1}{2} - \varepsilon) \right] \, ,
}
with
\SP{
\zeta_p (s) &\equiv \mu^{1+2s} \sum_{l=0}^{\infty} l (l+1) \left[ (l+\tfrac{1}{2})^2 + m^2 R^2 \right]^{-s} \, , \\
&= \mu^{1+2s} \sum_{l=0}^{\infty} \bigg[ \left[(l+\tfrac{1}{2})^2 + m^2 R^2 \right]^{1-s} - (\tfrac{1}{4} + m^2 R^2) \left[(l+\tfrac{1}{2})^2 + m^2 R^2 \right]^{-s} \bigg] \, .
}
Then, using (\ref{zeta_fermion}), the Casimir energy takes the $\mu$-dependent form
\SP{
E_{Cas}^{(f)}(\mu) = &- \frac{d_{{\cal R}}^{(f)}}{R} \Bigg{\{} m^2 R^2 \left[ \frac{1}{16} + \frac{1}{32} m^2 R^2 + \frac{1}{8} (1 + m^2 R^2) \log \left( \frac{m R}{2 \mu} \right) \right]\\
&+ \sum_{n=1}^{\infty} (-1)^n \left[ \frac{3}{2} \left( \frac{m R}{\pi n} \right)^2 K_{2} (2 \pi n m R) + (m^2 R^2 + \tfrac{1}{4}) \left( \frac{m R}{\pi n} \right) K_{1} (2 \pi n m R) \right] \Bigg{\}} \, .
\label{cas_sca_mu}
}
Equating this result with the cutoff-independent contributions in (\ref{cutoff-cas-final-fer}) gives the normalization factor $\mu \equiv e^{- \gamma_E}$, as in the scalar case. Therefore, the $\mu \equiv e^{- \gamma_E}$ scheme $I$-regularized fermion Casimir energy is
\SP{
E_{Cas}^{(f)I} = &- \frac{d_{{\cal R}}^{(f)}}{R} \Bigg{\{} m^2 R^2 \left[ \frac{1}{16} + \frac{1}{32} m^2 R^2 + \frac{1}{8} (1 + m^2 R^2) \gamma_E + \frac{1}{8} (1 + m^2 R^2) \log \left( \frac{m R}{2} \right) \right]\\
&+ \sum_{n=1}^{\infty} (-1)^n \left[ \frac{3}{2} \left( \frac{m R}{\pi n} \right)^2 K_{2} (2 \pi n m R) + (m^2 R^2 + \tfrac{1}{4}) \left( \frac{m R}{\pi n} \right) K_{1} (2 \pi n m R) \right] \Bigg{\}} \, .
\label{fercas1}
}
For regularization scheme $II$, where the Casimir energy is required to vanish as $m R \rightarrow \infty$, the normalization factor in (\ref{cas_sca_mu}) must be chosen as $\mu \equiv \frac{1}{2} m R \exp \left[ \frac{2 + m^2 R^2}{4(1 + m^2 R^2)} \right]$ such that the Casimir energy is reduced to
\SP{
E_{Cas}^{(f)II} = &- \frac{d_{{\cal R}}^{(f)}}{R} \sum_{n=1}^{\infty} (-1)^n \left[ \frac{3}{2} \left( \frac{m R}{\pi n} \right)^2 K_{2} (2 \pi n m R) + (m^2 R^2 + \tfrac{1}{4}) \left( \frac{m R}{\pi n} \right) K_{1} (2 \pi n m R) \right] \, .
\label{fercas2}
}
\subsection{Vectors}

For convenience we reproduce the Casimir Energy for massless vectors. It is
\EQ{
E_{Cas}^{(v)} = \frac{d_{{\cal A}}}{R} \sum_{l=0}^{\infty} l(l+2) (l+1) = \frac{d_{{\cal A}}}{R} \sum_{l=1}^{\infty} l (l^2 - 1) = \frac{d_{{\cal A}}}{R} \left[ \zeta(-3) - \zeta(-1) \right] = \frac{11 d_{{\cal A}}}{120 R} \, .
\label{vec_cas}
}
\section{High temperature partition functions}
\label{highTlogZ}

The high temperature partition functions for vectors, scalars, and fermions are obtained under the assumption that the theory is in the deconfined phase, that is, $\theta_i = 0$ and ${\mathscr P}^{({\cal R})}_n = \Tr_{{\cal R}} e^{i n \theta} = d_{{\cal R}}$. The range of temperatures for which this assumption is valid depends on the matter content of the theory which determines the existence and location of any critical temperatures at which the theory undergoes a phase transition. Analytical results obtained in the high temperature limit are compared to numerical results which are valid at all temperatures, while the theory remains in the deconfined phase.

\subsection{Massless vectors}

\begin{figure}[t]
  \hfill
  \begin{minipage}[t]{.49\textwidth}
    \begin{center}
\includegraphics[width=0.99\textwidth]{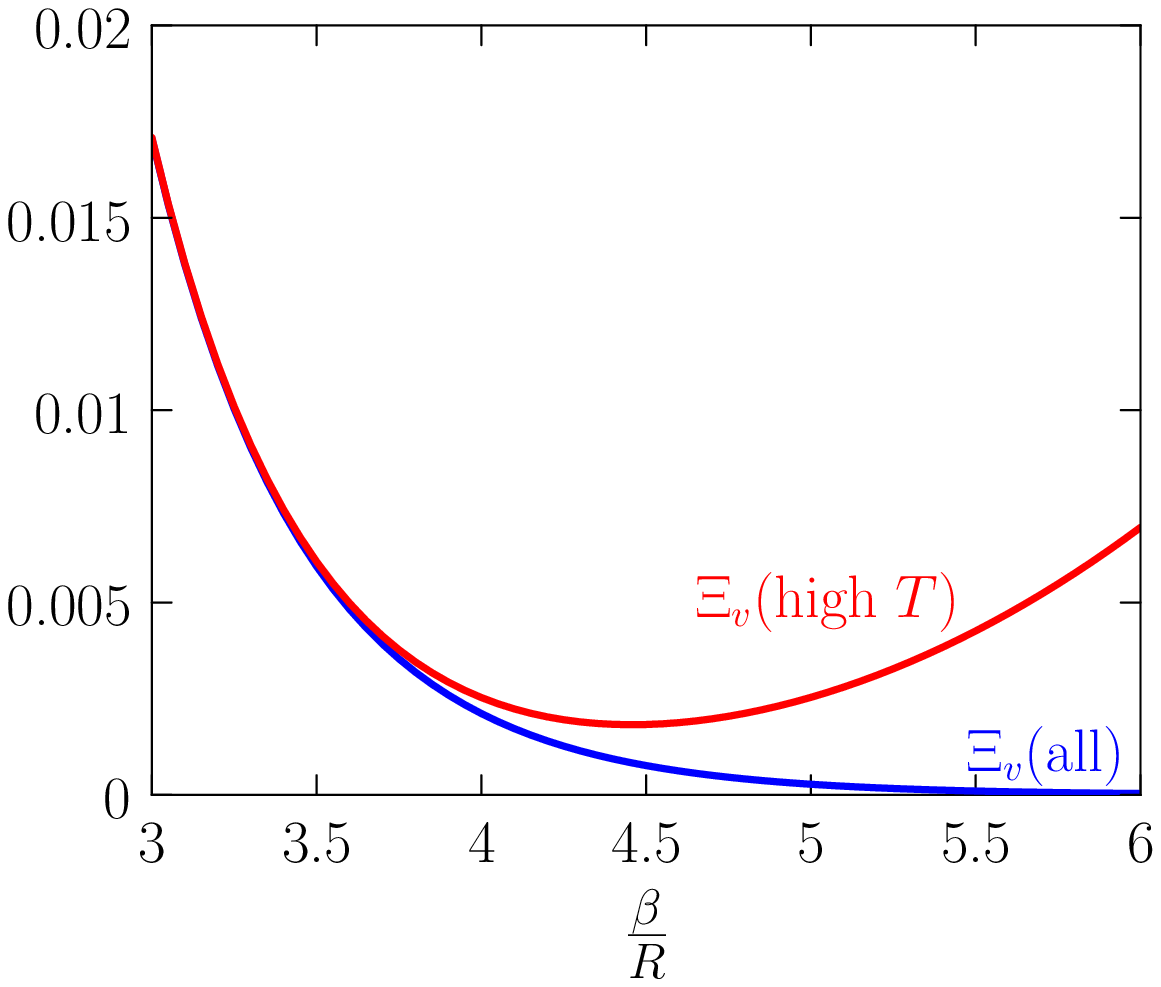}
    \end{center}
  \end{minipage}
  \hfill
  \begin{minipage}[t]{.49\textwidth}
    \begin{center}
\includegraphics[width=0.99\textwidth]{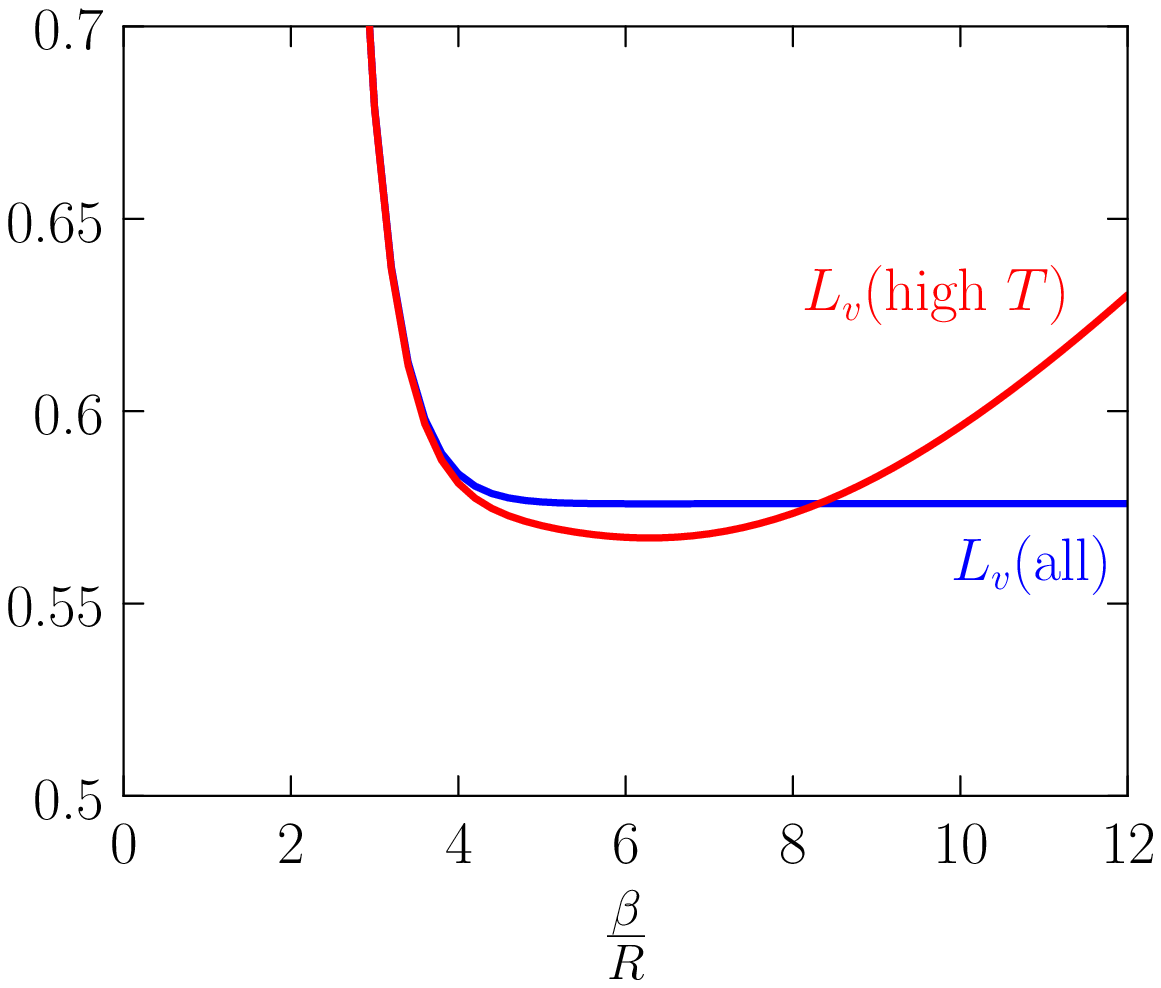}
    \end{center}
  \end{minipage}
  \hfill
\caption{(Left) Comparison of the full vector sum $\Xi_v ({\rm all})$ and the high temperature approximation $\Xi_v ({\rm high}~T)$ as a function of $\frac{\beta}{R}$. (Right) $L_v$ as a function of $\frac{\beta}{R}$ for theories with massless vectors in the deconfined phase using the full $\Xi_v$, labeled by $L_v ({\rm all})$, and the high temperature approximation, labeled by $L_v ({\rm high}~T)$.} 
\label{plot_Lvec_dec}
\end{figure}

First we consider the high temperature expansion of massless adjoint vectors. The partition function is obtained from
\EQ{
\log Z_v = \Xi_v - \beta E_{Cas}^{(v)} \, ,
\label{logZv}
}
where using (\ref{logZv1}), (\ref{eigenv}), (\ref{logdet}), and setting $\theta_i = 0$ gives
\EQ{
\Xi_{v} = 2 d_{{\cal A}} \sum_{n=1}^{\infty} \frac{1}{n} \sum_{l=0}^{\infty} l (l+2) e^{-n \beta (l+1)/R}\, .
\label{vec_sum_all}
}
Following \cite{Cardy:1991kr,Kutasov:2000td} the Mellin Transform (\ref{mellin}) is used to define the exponential in terms of a contour integral such that the sum takes the form
\EQ{
\Xi_{v} = \frac{2 d_{{\cal A}}}{2 \pi i} \int_{{\cal C}} {\rm d}s \, \Gamma(s) \left( \frac{\beta}{R} \right)^{-s} \zeta(s+1) \left[ \zeta(s-2) - \zeta(s) \right] \, ,
}
where ${\cal C}$ is the contour in the complex $s$-plane given by $c_0 - i \infty \rightarrow c_0 + i \infty$ with $c_0 > 3$. Adding an arc in the left-half $s$-plane to enclose the poles allows for the extraction of high temperature contributions to the sum using the residue theorem. The low-temperature corrections are contained in the remaining integral over the arc, which can be checked by computing it numerically (and perhaps it is even calculable using a similar approach as the one in \cite{Elizalde:1988xc}). It is sufficient for our purposes to drop this contribution (we will however compare the high temperature results with numerical results obtained by evaluating the sums in (\ref{vec_sum_all}) directly to show where the approximation holds and that our conclusions do not change by including the low temperature corrections). Performing the closed contour integral leads to the result 
\EQ{
\Xi_{v} \simeq d_{{\cal A}} \left[ \frac{2 \pi^4 R^3}{45 \beta^3} - \frac{\pi^2 R}{3 \beta} - \frac{1}{2 \pi^2} \zeta(3) + \log \left( \frac{2 \pi R}{\beta} \right) + \frac{11 \beta}{120 R} \right] \, .
\label{vec_sum_apr}
}
To see where this approximation begins to show small deviations from the full form in (\ref{vec_sum_all}) refer to Figure \ref{plot_Lvec_dec} (Left) which compares them as a function of $\frac{\beta}{R}$. Note the range of $\frac{\beta}{R}$ is significant compared to the scale of the deviations in $\Xi_v$.

Adding the Casimir contribution $- \beta E_{Cas}^{(v)} = - d_{{\cal A}} \frac{11 \beta}{120 R}$ from (\ref{vec_cas}), the partition function (\ref{logZv}) is
\EQ{
\log Z_v \simeq d_{{\cal A}} \left[ \frac{2 \pi^4 R^3}{45 \beta^3} - \frac{\pi^2 R}{3 \beta} - \frac{1}{2 \pi^2} \zeta(3) + \log \left( \frac{2 \pi R}{\beta} \right) \right] \, .
}
To check if the Bekenstein bound is satisfied we solve for $L$ by plugging the partition function into (\ref{Lbek}) to obtain
\EQ{
L_v \simeq d_{{\cal A}} \bigg[ \frac{4 \pi^5 R^4}{15 \beta^4} - \frac{8 \pi^4 R^3}{45 \beta^3} - \frac{2 \pi^3 R^2}{3 \beta^2} + \left( \frac{\pi}{3} + 1\right) \frac{2 \pi R}{\beta} - 1 + \frac{1}{2 \pi^2} \zeta(3) + \log \left( \frac{\beta}{2 \pi R} \right) \bigg] \, .
\label{Lv_apr}
}
Figure \ref{plot_Lvec_dec} (Right) shows $L_v$ as a function of $\frac{\beta}{R}$ using the high temperature approximation in (\ref{Lv_apr}) and compares it with the numerical calculation for $L_v$, labeled $L_v ({\rm all})$, including the low temperature contributions in (\ref{vec_sum_all}). In either case $L_v > 0$ for all $\frac{\beta}{R}$. It is clear that $L_v ({\rm all})$ quickly reaches its asymptotic value $L_v ({\rm all}) \xrightarrow[\beta \rightarrow \infty]{} 2 \pi R E_{Cas}^{(v)} = \frac{11 \pi d_{{\cal A}}}{60} \approx 0.576 d_{{\cal A}}$. The minimum of $L_v$ in (\ref{Lv_apr}) is
\EQ{
L_v \Big|_{\beta = 2 \pi R} \simeq d_{{\cal A}} \left( \frac{29 \pi}{180} + \frac{1}{2 \pi^2} \zeta(3) \right) \approx 0.567 d_{{\cal A}} \, .
}
\subsection{Scalars}

The partition function for real scalars of mass $m$ and in the representation ${\cal R}$ is obtained from
\EQ{
\log Z_s = N_s (\Xi_s - \beta E_{Cas}^{(s)}) \, ,
\label{logZs}
}
where using (\ref{logZs1}), (\ref{eigens}), (\ref{logdet}), and setting $\theta_i = 0$ gives
\EQ{
\Xi_s = d_{{\cal R}} \sum_{n=1}^{\infty} \frac{1}{n} \sum_{l=0}^{\infty} (l+1)^2 e^{-n \frac{\beta}{R} \sqrt{(l+1)^2 + m^2 R^2}} \, .
\label{sca_sum_all}
}
As in the case for vectors we consider the exponential as a contour integral such that $\Xi_s$ takes the form
\EQ{
\Xi_s = \frac{d_{{\cal R}}}{2 \pi i} \int_{{\cal C}} {\rm d}s \, G_s (s) \, ,
\label{scalar_contour_integral}
}
with
\SP{
G_s (s) \equiv &\zeta(s+1) \Gamma(s) \left( \frac{\beta}{R} \right)^{-s} \sum_{l=1}^{\infty} l^2 \left[ l^2 + m^2 R^2 \right]^{-s/2} \, ,\\
= &\zeta(s+1) \Gamma(s) \left( \frac{\beta}{R} \right)^{-s} \sum_{l=1}^{\infty} \Bigg[ \left[ l^2 + m^2 R^2 \right]^{1-s/2} - m^2 R^2 \left[ l^2 + m^2 R^2 \right]^{-s/2} \Bigg] \, .
\label{Gs_sca}
}
\subsubsection{Massless scalars}

\begin{figure}[t]
  \hfill
  \begin{minipage}[t]{.49\textwidth}
    \begin{center}
\includegraphics[width=0.99\textwidth]{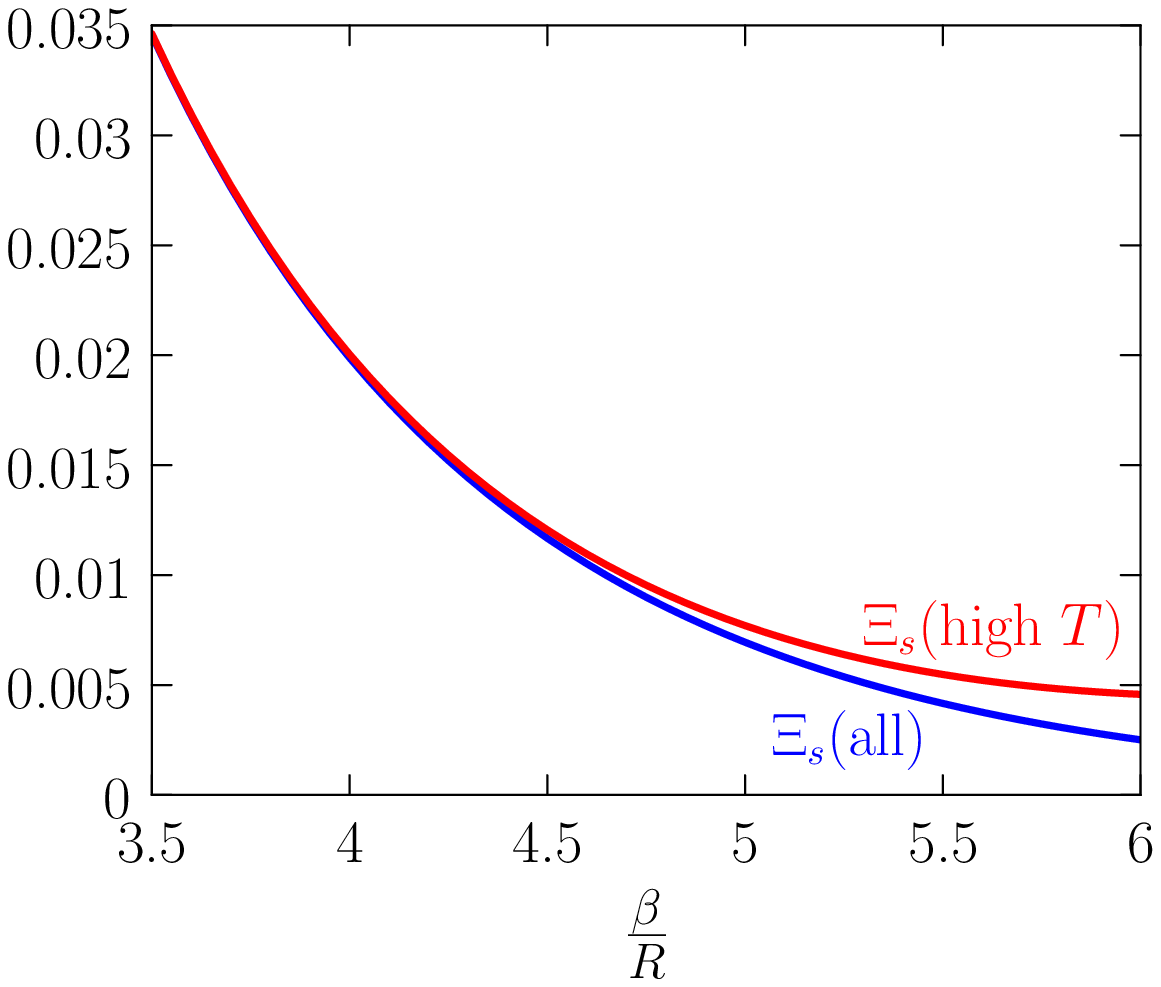}
    \end{center}
  \end{minipage}
  \hfill
  \begin{minipage}[t]{.49\textwidth}
    \begin{center}
\includegraphics[width=0.99\textwidth]{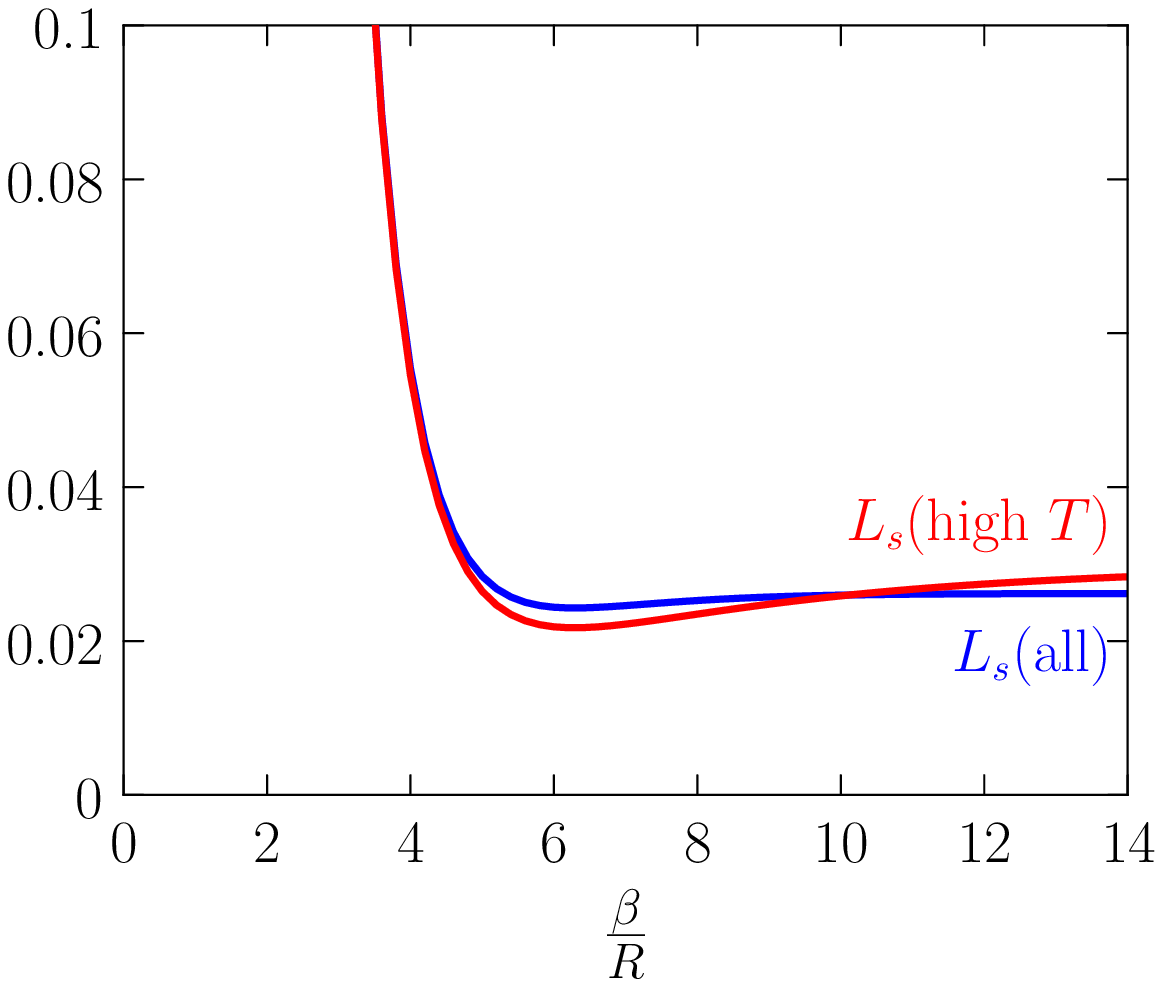}
    \end{center}
  \end{minipage}
  \hfill
\caption{(Left) Comparison of the full scalar sum $\Xi_s ({\rm all})$ and the high temperature approximation $\Xi_s ({\rm high}~T)$ as a function of $\frac{\beta}{R}$. (Right) $L_s$ as a function of $\frac{\beta}{R}$ for theories with massless scalars in the deconfined phase using the full $\Xi_s$, labeled $L_s ({\rm all})$, and the high temperature approximation, labeled $L_s ({\rm high}~T)$.} 
\label{plot_Lsca0}
\end{figure}

In the massless limit (\ref{Gs_sca}) reduces to
\EQ{
G_s (s) = \zeta(s+1) \Gamma(s) \left( \frac{\beta}{R} \right)^{-s} \zeta(s-2) \, .
}
The integral in (\ref{scalar_contour_integral}) is calculated by adding an arc in the left-half $s$-plane to enclose the poles. Then collecting the residues from the closed contour integral gives the high temperature result
\EQ{
\Xi_s \simeq d_{{\cal R}} \left[ \frac{\pi^4 R^3}{45 \beta^3} - \frac{1}{4 \pi^2} \zeta(3) + \frac{\beta}{240 R} \right] \, .
\label{scaxi0}
}
Figure \ref{plot_Lsca0} (Left) compares this result against the full form of $\Xi_s$ in (\ref{sca_sum_all}) evaluated numerically. The partition function is obtained by plugging (\ref{scaxi0}) into (\ref{logZs}) and including the Casimir energy in (\ref{scacas0}) which results in
\EQ{
\log Z_s \simeq N_s d_{{\cal R}} \left[ \frac{\pi^4 R^3}{45 \beta^3} - \frac{1}{4 \pi^2} \zeta(3) \right] \, .
}
Using this in (\ref{Lbek}) gives
\EQ{
L_s \simeq N_s d_{{\cal R}} \bigg[ \frac{2 \pi^5 R^4}{15 \beta^4} - \frac{4 \pi^4 R^3}{45 \beta^3} + \frac{1}{4 \pi^2} \zeta(3) \bigg] \, ,
\label{sca_L_approx}
}
which is positive for all $\frac{\beta}{R}$. Figure \ref{plot_Lsca0} (Right) shows $L_s$ as a function of $\frac{\beta}{R}$ and compares with the full numerical result, $L_s({\rm all})$, which includes the low temperature contributions by performing the sums in (\ref{sca_sum_all}) directly. It appears that $L_s ({\rm all}) > 0$ for all $\frac{\beta}{R}$ as it quickly approaches the asymptotic value $L_s({\rm all}) \xrightarrow[\beta \rightarrow \infty]{} 2 \pi R E_{Cas}^{(s)} = \frac{\pi d_{{\cal R}}^{(s)}}{120} \approx 0.0262 d_{{\cal R}}^{(s)}$. The minimum of $L_s$ in (\ref{sca_L_approx}) is
\EQ{
L_s \Big|_{\beta = 2 \pi R} \simeq d_{{\cal R}} N_s \left[ - \frac{\pi}{360} + \frac{1}{4 \pi^2} \zeta(3) \right] \approx 0.0217 ~ d_{{\cal R}} N_s \, .
}
\subsubsection{Massive scalars}

Using (\ref{zeta_scalar}) to perform the sum over $l$, $G_s$ simplifies to
\SP{
G_s &(s) = 2^{s} \zeta(s+1) \Gamma(\tfrac{s+1}{2}) \left( \frac{\beta}{R} \right)^{-s} \bigg[ \frac{1}{8} (m R)^{3-s} \Gamma(\tfrac{s-3}{2})\\
&+ \sum_{n=1}^{\infty} \bigg[ (\tfrac{s}{2} - 1) \left( \frac{m R}{\pi n} \right)^{\frac{3-s}{2}} K_{\frac{s-3}{2}} (2 \pi n m R) - m^2 R^2 \left( \frac{m R}{\pi n} \right)^{\frac{1-s}{2}} K_{\frac{s-1}{2}} (2 \pi n m R) \bigg] \bigg] \, .
}
The Gamma functions and zeta function in $G_s(s)$ suggest that poles are possible for $s = 3, 1, 0, -1, -3, -5, ...$. In fact $G_s(s)$ has non-zero residue for all of these values such that the closed contour integral gives
\SP{
\Xi_s \simeq &d_{{\cal R}} \Bigg{\{} \frac{\pi^4 R^3}{45 \beta^3} - \frac{m^2 \pi^2 R^3}{12 \beta} + \frac{1}{6} \pi m^3 R^3 + \frac{1}{2} \bigg[ m^2 R^2 \log \left( 1 - e^{-2 \pi m R} \right)\\
&- \frac{m R}{\pi} {\rm Li}_2 (e^{-2 \pi m R}) - \frac{1}{2 \pi^2} {\rm Li}_3 (e^{-2 \pi m R}) \bigg] + \frac{\beta}{2 R} m^4 R^4 \left[ -\frac{3}{32} + \frac{1}{8} \gamma_E + \frac{1}{8} \log \left( \frac{m \beta}{4 \pi} \right) \right]\\
&+ \frac{\beta}{2 R} \sum_{n=1}^{\infty} \left[ \frac{3}{2} \left( \frac{m R}{\pi n} \right)^2 K_2 (2 \pi n m R) + m^2 R^2 \left( \frac{m R}{\pi n} \right) K_1 (2 \pi n m R) \right]\\
&+ \xi_s (\beta/R, m R) \Bigg{\}} \, ,
\label{massive_sca_sum}
}
where $\xi_s (\beta/R, m R) = {\cal O}\left( \frac{\beta}{R} \right)^3 {\cal O} (m R)^6$ and is given by
\EQ{
\xi_s (\beta/R, m R) = \frac{m^4 R^3 \beta}{8} \sum_{n=1}^{\infty} (-1)^n \left( \frac{m \beta}{4 \pi} \right)^{2 n} \frac{\Gamma(2 n + 1) \zeta(2 n + 1)}{\Gamma(n+1) \Gamma(n+3)} \, .
\label{xi_mass}
}
The third line in (\ref{massive_sca_sum}) is always canceled by the Casimir contribution, and, if regularization scheme I is used to obtain the Casimir energy, where $\mu \equiv e^{-\gamma_E}$, then part of the second line is cancelled off as well. The partition function, (\ref{logZs}), is then obtained using the scheme $I$-regularized Casimir energy in (\ref{scacas1}), resulting in
\SP{
\log Z_s^{I} \simeq &d_{{\cal R}} \Bigg{\{} \frac{\pi^4 R^3}{45 \beta^3} - \frac{m^2 \pi^2 R^3}{12 \beta} + \frac{1}{6} \pi m^3 R^3 + \frac{1}{2} \bigg[ m^2 R^2 \log \left( 1 - e^{-2 \pi m R} \right)\\
&- \frac{m R}{\pi} {\rm Li}_2 (e^{-2 \pi m R}) - \frac{1}{2 \pi^2} {\rm Li}_3 (e^{-2 \pi m R}) \bigg] + \frac{\beta}{16 R} m^4 R^4 \left[ -1 + \log \left( \frac{\beta}{2 \pi R} \right) \right]\\
&+ \xi_s (\beta/R, m R) \Bigg{\}} \, .
}
It is now possible to consider whether the Bekenstein bound is satisfied for theories with massive fundamental scalars. From (\ref{Lbek}) $L$ takes the form
\SP{
L_s^{I} \simeq &N_s d_{{\cal R}} \bigg[ \frac{2 \pi^5 R^4}{15 \beta^4} - \frac{4 \pi^4 R^3}{45 \beta^3} - \frac{\pi^3 m^2 R^4}{6 \beta^2} + \frac{\pi^2 m^2 R^3}{6 \beta} - \frac{1}{6} \pi m^3 R^3\\
&- \frac{1}{2} m^2 R^2 \log \left( 1 - e^{-2 \pi m R} \right) + \frac{m R}{2 \pi} {\rm Li}_2 (e^{-2 \pi m R}) + \frac{1}{4 \pi^2} {\rm Li}_3 (e^{-2 \pi m R}) + \frac{1}{16} m^4 R^3 \beta\\
&+ \frac{1}{8} \pi m^4 R^4 \log \left( \frac{2 \pi R}{\beta} \right) + \xi_s' (\beta/R, m R) \bigg] \, ,
\label{sca_L_approx_m1}
}
where $\xi_s' (\beta/R, m R) = {\cal O}\left( \frac{\beta}{R} \right)^2 {\cal O} (m R)^6$ and is given by
\EQ{
\xi_s' (\beta/R, m R) = \frac{m^4 R^3}{4} \sum_{n=1}^{\infty} (-1)^n \left( \frac{m \beta}{4 \pi} \right)^{2 n} \frac{\Gamma(2 n + 1)\zeta(2 n + 1)}{\Gamma(n+1) \Gamma(n+3)} \left[ (\beta - 2 \pi R) n - \pi R \right] \, .
\label{xi_mass}
}
To determine the behavior of $L_s$ for $m R \ne 0$ it is necessary to determine if the sum in $\xi_s' (\beta/R, m R)$ converges. Consider the large $n$ limit. Since $\lim_{n \rightarrow \infty} \frac{\Gamma(2n+1)}{\Gamma(n+1)\Gamma(n+3)} = \frac{2^{2n}}{n^{5/2} \sqrt{\pi}}$ the limit of $\xi_s'$ is
\EQ{
\lim_{n \rightarrow \infty} \xi_s' (\beta / R, m R) = \frac{m^4 R^3}{4 n^{5/2} \sqrt{\pi}} (-1)^n \left( \frac{m \beta}{2 \pi} \right)^{2n} \left[ \left( \beta - 2 \pi R \right) n - \pi R \right] \, .
}
Setting aside the constants it turns out that convergence is determined by the limits
\SP{
\frac{(-1)^n}{n^{5/2}} \left( \frac{m \beta}{2 \pi} \right)^{2n} \left[ \left( \beta - 2 \pi R \right) n - \pi R \right] &\xrightarrow[n \rightarrow \infty]{} \hspace{1cm} 0, \hspace{1cm} {\rm if}~ m \beta \le 2\pi \, ,\\
&\xrightarrow[n \rightarrow \infty]{} \hspace{5mm} \pm \infty, \hspace{1cm} {\rm if}~ m \beta > 2\pi \, .
}
Therefore, it is possible to approximate $L_s$ for $m \beta \le 2 \pi$ by truncating the sum over $n$. However, for $m \beta > 2 \pi$ the result will go to $\pm \infty$, depending on whether the sum is truncated at even or odd $n$, unless $\xi_s'(\beta/R, m R)$ can be regularized.

\begin{figure}[t]
  \hfill
  \begin{minipage}[t]{.49\textwidth}
    \begin{center}
\includegraphics[width=0.99\textwidth]{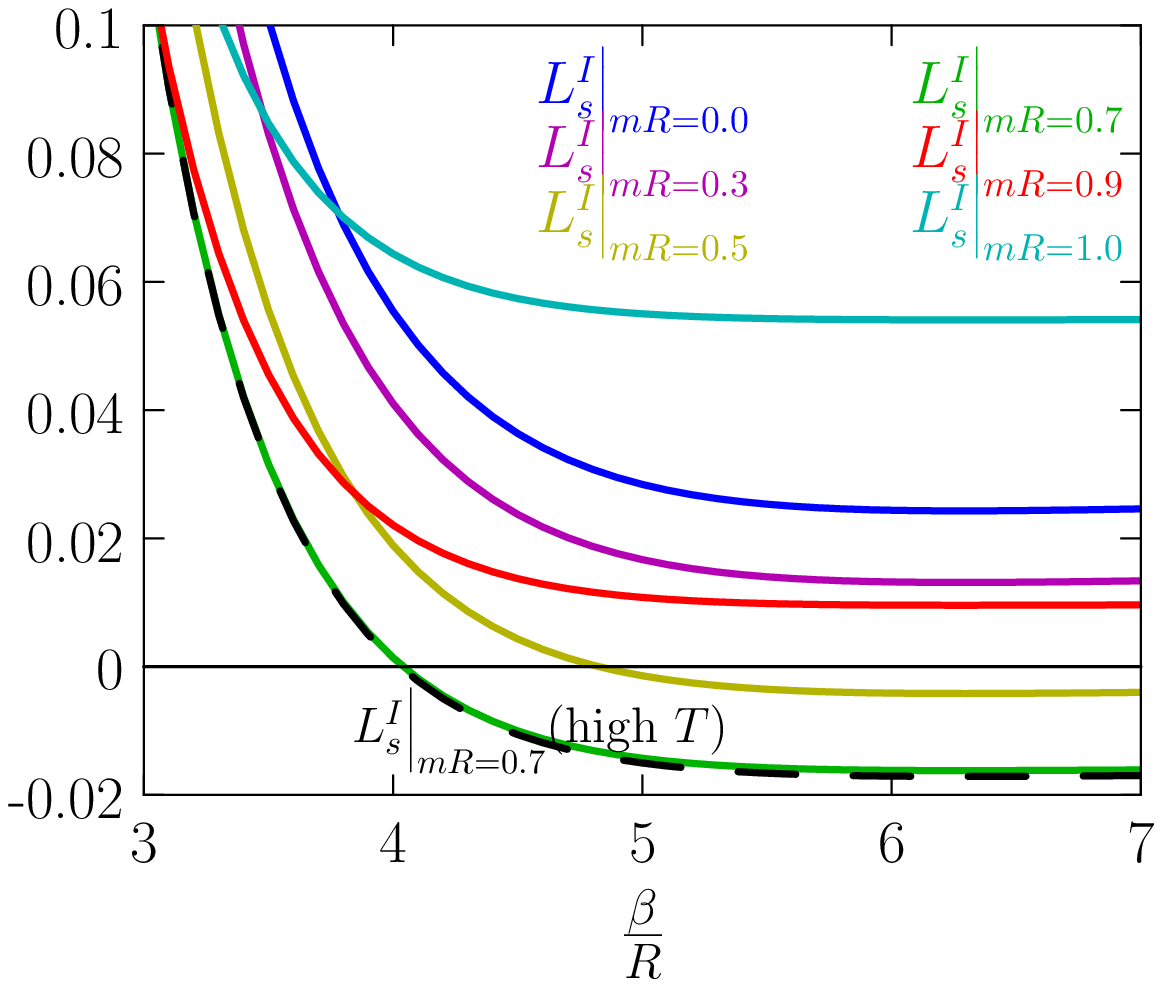}
    \end{center}
  \end{minipage}
  \hfill
  \begin{minipage}[t]{.49\textwidth}
    \begin{center}
\includegraphics[width=0.99\textwidth]{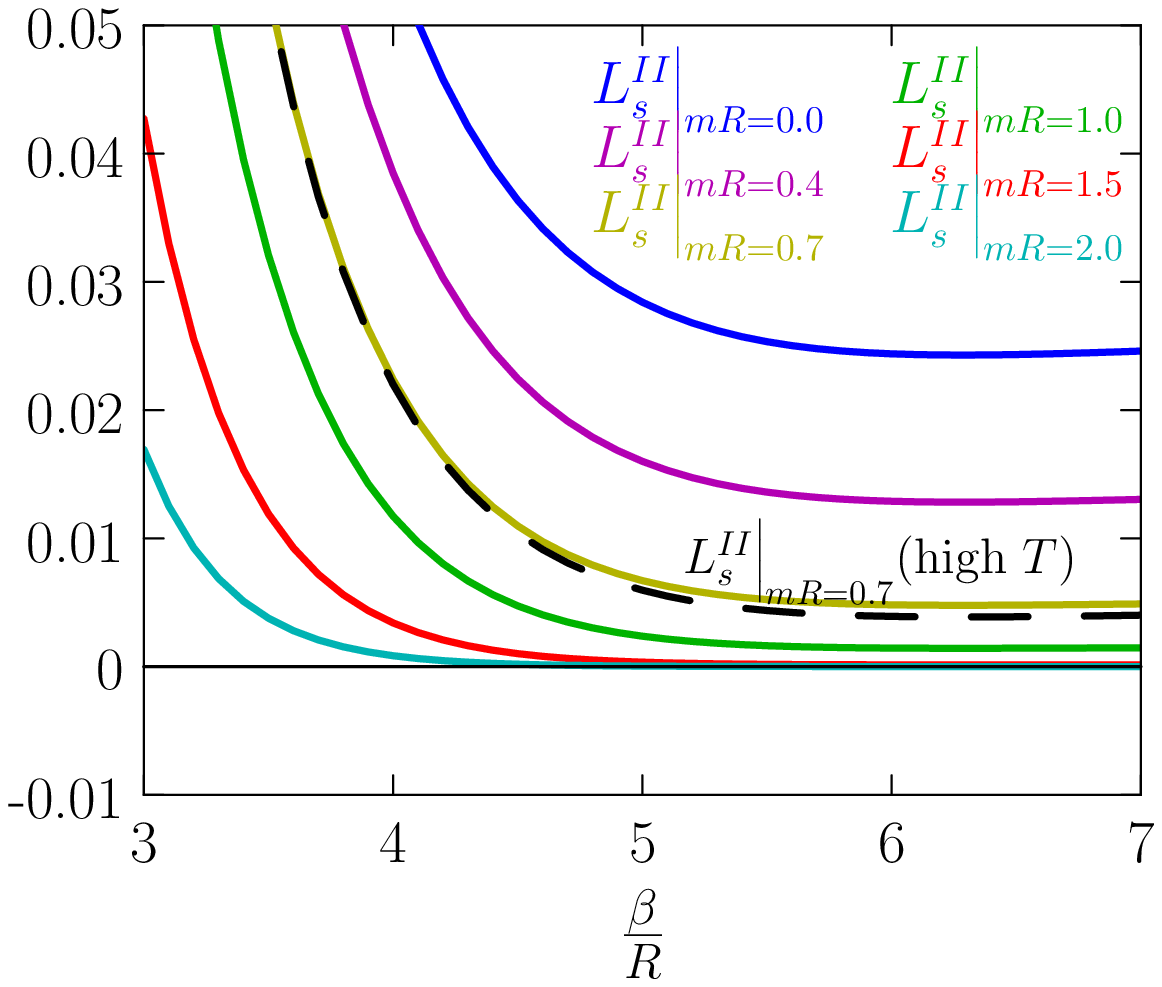}
    \end{center}
  \end{minipage}
  \hfill
\caption{$L$ for scalars with various masses as a function of $\frac{\beta}{R}$ including low temperature contributions. (Left) $L_s$ calculated with scheme $I$-regularized Casimir energy where $\mu \equiv e^{-\gamma_E}$. (Right) $L_s$ calculated with scheme $II$-regularized Casimir energy where $\mu \equiv \frac{1}{2} m R \exp \left[ \frac{2 + m^2 R^2}{4(1+m^2 R^2)} \right]$. The dotted line corresponds to the high $T$ approximation of $L_s$ for $m R = 0.7$.} 
\label{plot_Lsm}
\end{figure}

Figure \ref{plot_Lsm} (Left) shows $L_s^I$ in (\ref{sca_L_approx_m1}), referred to as $L_s^I ({\rm high}~T)$, as a function of $\frac{\beta}{R}$ for $m R = 0.7$ along with numerical calculations of $L_s^I$, including the low temperature contributions in (\ref{sca_sum_all}), for various $m R$. It turns out that when regularization scheme $I$ is used to obtain the Casimir energy, there exists a critical range $0.455 \lapprox m R \lapprox 0.864$ within which $L_s^{I} < 0$ for a range of $\frac{\beta}{R}$.

If regularization scheme II is used instead, with $\mu \equiv \frac{1}{2} m R \exp \left[ \frac{2 + m^2 R^2}{4(1+m^2 R^2)} \right]$, then the Casimir energy in (\ref{scacas2}) is used to obtain the partition function from (\ref{logZs}) such that
\SP{
\log Z_s^{II} \simeq &d_{{\cal R}} \Bigg{\{} \frac{\pi^4 R^3}{45 \beta^3} - \frac{m^2 \pi^2 R^3}{12 \beta} + \frac{1}{6} \pi m^3 R^3 + \frac{1}{2} \bigg[ m^2 R^2 \log \left( 1 - e^{-2 \pi m R} \right)\\
&- \frac{m R}{\pi} {\rm Li}_2 (e^{-2 \pi m R}) - \frac{1}{2 \pi^2} {\rm Li}_3 (e^{-2 \pi m R}) \bigg]\\
&+ \frac{\beta}{16 R} m^4 R^4 \left[ -\frac{3}{4} + \gamma_E + \log \left( \frac{m \beta}{4 \pi} \right) \right] + \xi_s (\beta/R, m R) \Bigg{\}} \, .
}
Using this in (\ref{Lbek}) gives
\SP{
L_s^{II} \simeq &N_s d_{{\cal R}} \bigg[ \frac{2 \pi^5 R^4}{15 \beta^4} - \frac{4 \pi^4 R^3}{45 \beta^3} - \frac{\pi^3 m^2 R^4}{6 \beta^2} + \frac{\pi^2 m^2 R^3}{6 \beta} - \frac{1}{6} \pi m^3 R^3 - \frac{1}{8} \pi m^4 R^4 (\tfrac{1}{4} + \gamma_E)\\
&- \frac{1}{2} m^2 R^2 \log \left( 1 - e^{-2 \pi m R} \right) + \frac{m R}{2 \pi} {\rm Li}_2 (e^{-2 \pi m R}) + \frac{1}{4 \pi^2} {\rm Li}_3 (e^{-2 \pi m R}) + \frac{1}{16} m^4 R^3 \beta\\
&+ \frac{1}{8} \pi m^4 R^4 \log \left( \frac{4 \pi}{m \beta} \right) + \xi_s' (\beta/R, m R) \bigg] \, .
\label{sca_L_approx_m2}
}
The result is plotted in Figure \ref{plot_Lsm} (Right) as a function of $\frac{\beta}{R}$ for $m R = 0.7$ along with numerical calculations of the full result for various $m R$ including the low temperature contributions in (\ref{sca_sum_all}). In contrast to $L_s^I$ it appears that $L_s^{II} > 0$ for all $\frac{\beta}{R}$ regardless of the value of $m R$. The difference between $L_s^{II}$ and $L_s^I$ is
\EQ{
L_s^{II} - L_s^{I} \simeq \frac{\pi m^4 R^4}{8} \left[ - \frac{1}{4} - \gamma_E + \log 2 - \log \left( m R \right) \right] \, ,
}
which is positive for $m R < 2 e^{-\frac{1}{4} - \gamma_E} \simeq 0.875$, after which $L^{II} < L^{I}$. This suggests that $L_s^I$ is only able to violate the Bekenstein bound when $m R \lapprox 0.875$, which is satisfied by the range determined numerically, $0.455 \lapprox m R \lapprox 0.864$.
\subsection{Fermions}

The contribution to the partition function of $N_f$ Weyl (Majorana) fermions of mass $m$ in representation ${\cal R}$ is given by
\EQ{
\log Z_f = -N_f (\Xi_f + \beta E_{Cas}^{(f)}) \, ,
\label{logZf}
}
where using (\ref{logZf1}), (\ref{eigenf}), (\ref{logdet}), and setting $\theta_i = 0$ gives
\EQ{
\Xi_f = 2 d_{{\cal R}} \sum_{n=1}^{\infty} \frac{(-1)^n}{n} \sum_{l=1}^{\infty} l (l+1) e^{-n \frac{\beta}{R} \sqrt{(l+\frac{1}{2})^2 + m^2 R^2}} \, .
\label{fer_sum_all}
}
Applying the Mellin transform (\ref{mellin}) the sum takes the form
\EQ{
\Xi_f = \frac{2 d_{{\cal R}}}{2 \pi i} \int_{{\cal C}} {\rm d}s \, G_f (s) \, ,
\label{fermion_contour_integral}
}
where the integrand is given by
\SP{
G_f (s) \equiv & \Gamma(s) \left( \frac{\beta}{R} \right)^{-s} \sum_{n=1}^{\infty} \frac{(-1)^n}{n^{s+1}} \sum_{l=0}^{\infty} l (l+1) \left[ \left(l+\tfrac{1}{2}\right)^2 + m^2 R^2 \right]^{-s/2} \, ,\\
= & (-1+2^{-s}) \zeta(s+1) \Gamma(s) \left( \frac{\beta}{R} \right)^{-s} \sum_{l=0}^{\infty} \Bigg[ \left[ \left(l+\tfrac{1}{2}\right)^2 + m^2 R^2 \right]^{1-s/2}\\
&- \left( \tfrac{1}{4} + m^2 R^2 \right) \left[ \left(l+\tfrac{1}{2}\right)^2 + m^2 R^2 \right]^{-s/2} \Bigg] \, .
\label{Gf}
}
\subsubsection{Massless fermions}

\begin{figure}[t]
  \hfill
  \begin{minipage}[t]{.49\textwidth}
    \begin{center}
\includegraphics[width=0.99\textwidth]{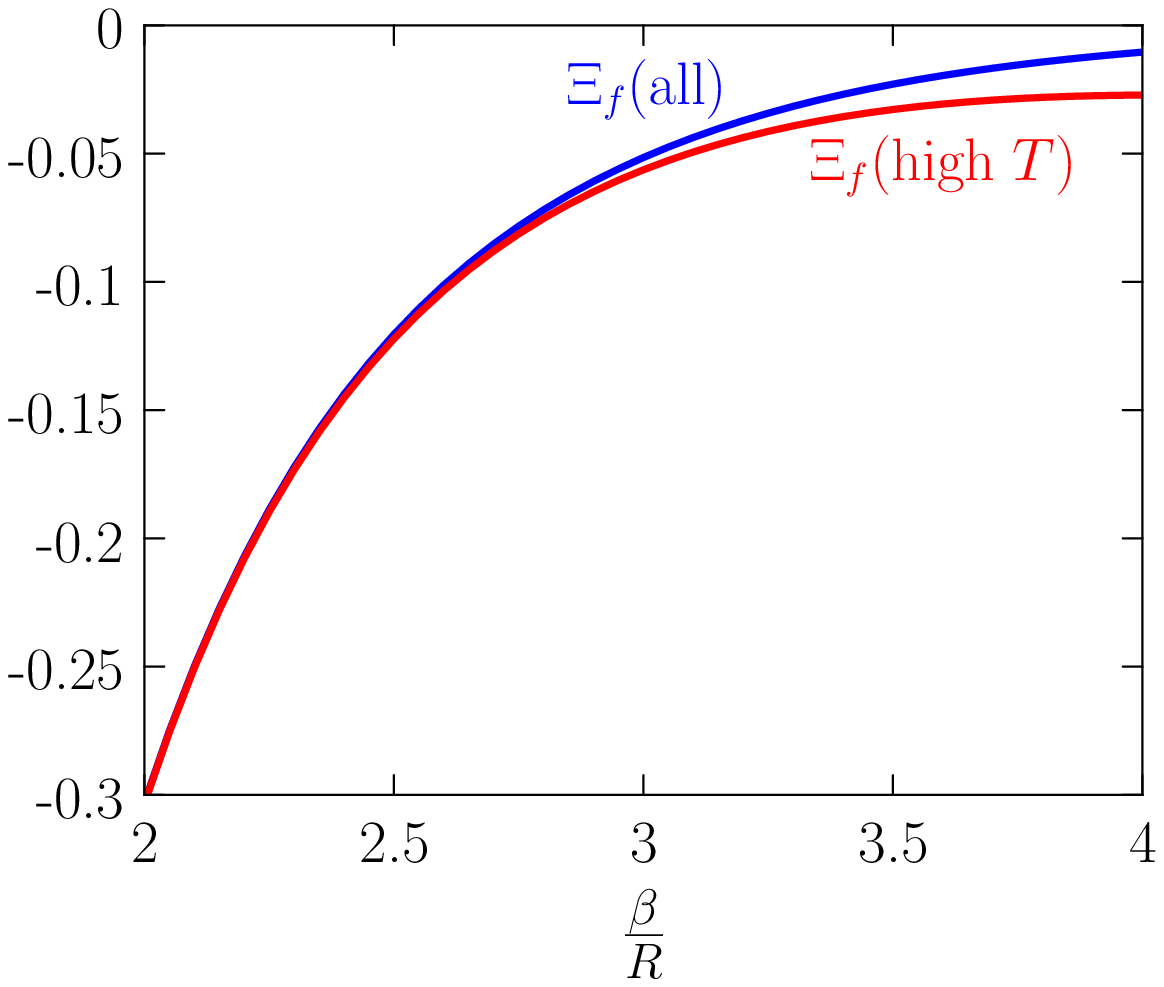}
    \end{center}
  \end{minipage}
  \hfill
  \begin{minipage}[t]{.49\textwidth}
    \begin{center}
\includegraphics[width=0.99\textwidth]{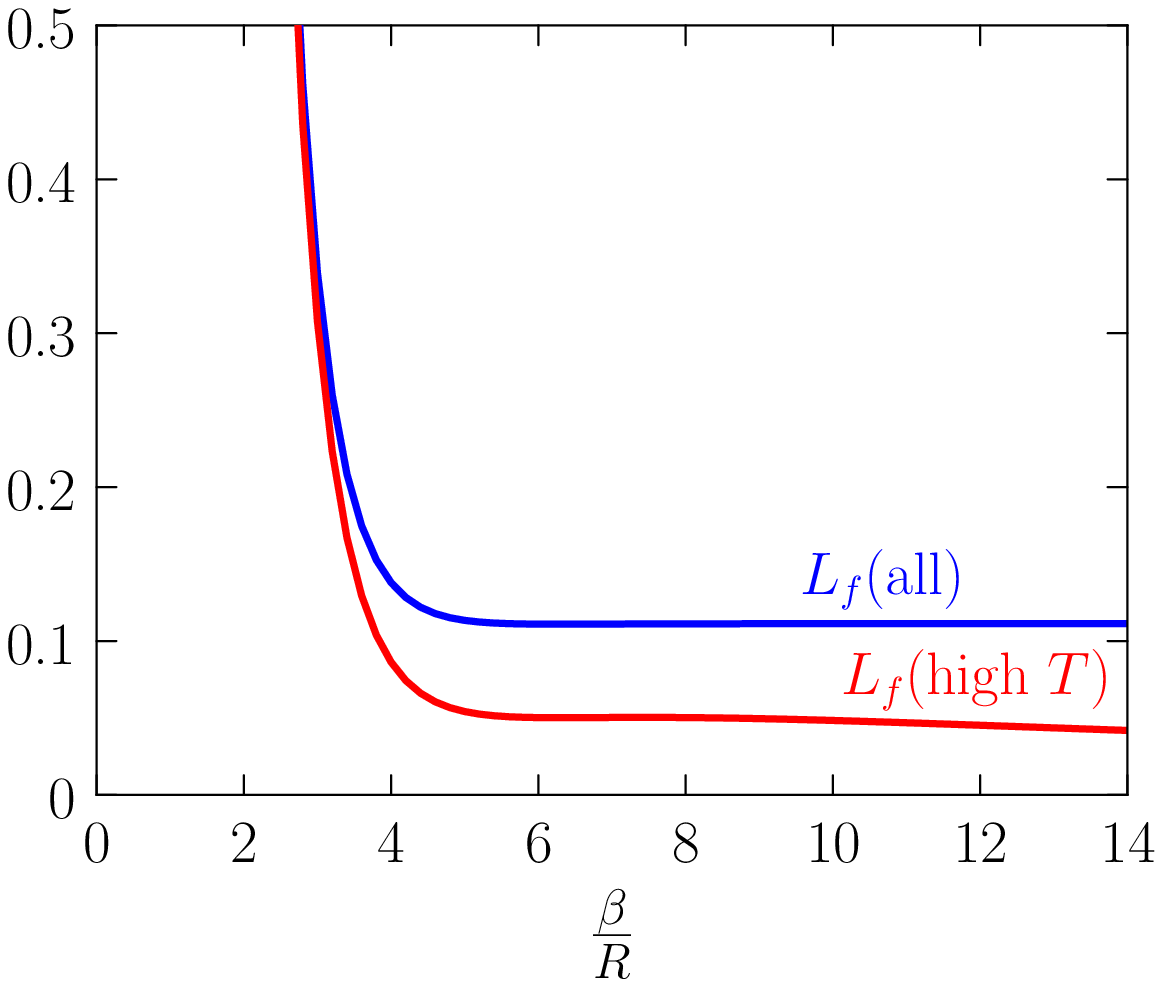}
    \end{center}
  \end{minipage}
  \hfill
\caption{(Left) Comparison of the full $\Xi_f ({\rm all})$ and the high temperature approximation $\Xi_f ({\rm high}~T)$ as a function of $\frac{\beta}{R}$. (Right) $L_f$ as a function for $\frac{\beta}{R}$ for theories with massless fermions in the deconfined phase using the full $\Xi_f$, labeled $L_f ({\rm all})$, and the high temperature approximation, labeled $L_f ({\rm high}~T)$.} 
\label{plot_Lf0}
\end{figure}

In the massless limit the integrand simplies to
\EQ{
G_f(s) = (-1+2^{-s}) \zeta(s+1) \Gamma(s) \left( \frac{\beta}{R} \right)^{-s} \left[ \zeta(s-2, \tfrac{1}{2}) - \frac{1}{4} \zeta(s, \tfrac{1}{2}) \right] \, .
}
Collecting the non-zero residues at $s = 3, 1, -1$, the high temperature contribution from the closed contour integral takes the form
\SP{
\Xi_f \simeq &2 d_{{\cal R}} \left[ - \frac{7 \pi^4 R^3}{360 \beta^3} + \frac{\pi^2 R}{48 \beta} - \frac{17 \beta}{1920 R} \right] \, .
\label{ferxi0}
}
This result is plotted in Figure \ref{plot_Lf0} (Left) as a function of $\frac{\beta}{R}$, where it is referred to as $\Xi_f ({\rm high}~T)$, along with the direct numerical calculation of (\ref{fer_sum_all}), referred to as $\Xi_f ({\rm all})$, which includes the low temperature contributions. The partition function calculated from (\ref{logZf}) using (\ref{ferxi0}) and the Casimir energy in (\ref{fercas0}) is
\SP{
\log Z_f \simeq &N_f d_{{\cal R}} \left[ \frac{7 \pi^4 R^3}{180 \beta^3} - \frac{\pi^2 R}{24 \beta} \right] \, .
}
Using this in (\ref{Lbek}) gives
\SP{
L_f \simeq &N_f d_{{\cal R}} \bigg[ \frac{7 \pi^5 R^4}{30 \beta^4} - \frac{7 \pi^4 R^3}{45 \beta^3} + \frac{1}{12} \left( - \frac{\pi^3 R^2}{\beta^2} + \frac{\pi^2 R}{\beta} \right) \bigg] \, ,
\label{fer_L}
}
which is positive for all $\frac{\beta}{R}$ and is plotted in Figure \ref{plot_Lf0} (Right) and labeled $L_f ({\rm high}~T)$, along with the numerical calculation of $L_f$, labeled $L_f({\rm all})$, which includes low temperature contributions in (\ref{fer_sum_all}). It appears that $L_f ({\rm all}) > 0$ for all $\frac{\beta}{R}$ since it asymptotes quickly to $L_f ({\rm all}) \xrightarrow[\beta \rightarrow \infty]{} 2 \pi R E_{Cas}^{(f)} = \frac{17 \pi}{480} d_{{\cal R}}^{(f)} \approx 0.111 d_{{\cal R}}^{(f)}$. The minimum of $L_f$ in (\ref{fer_L}) is
\EQ{
L_f \Big|_{\frac{\beta}{R} = \infty} \simeq 0 \, .
}
\subsubsection{Massive fermions}

The sums over $l$ in (\ref{Gf}) can be performed using (\ref{zeta_fermion}) such that $G_f(s)$ simplifies to
\SP{
G_f (s) = & (1-2^{s}) \left( \frac{\beta}{R} \right)^{-s} \zeta(s+1) \Gamma(\tfrac{s+1}{2}) \bigg{\{} \frac{1}{8} (m R)^{1-s} \left( m^2 R^2 - \frac{1}{4} (s - 3) \right) \Gamma(\tfrac{s-3}{2})\\
&+ \sum_{n=1}^{\infty} (-1)^n \bigg[ (\tfrac{s}{2} - 1) \left( \frac{\pi n}{m R} \right)^{(s-3)/2} K_{\frac{s-3}{2}} (2 \pi n m R)\\
&\hspace{24mm} - (\tfrac{1}{4} + m^2 R^2) \left( \frac{\pi n}{m R} \right)^{(s-1)/2} K_{\frac{s-1}{2}} (2 \pi n m R) \bigg] \bigg{\}} \, .
}
Adding an arc at infinity in the left half complex $s$-plane to close the contour in (\ref{fermion_contour_integral}) around the poles in $G_f(s)$ results in non-zero residues for $s = 3, 1, -1, -3, -5, ...$. Collecting these results in
\SP{
\Xi_f \simeq &\, 2 d_{{\cal R}} \bigg{\{} - \frac{7 \pi^4 R^3}{360 \beta^3} + \frac{\pi^2 R}{24 \beta} \left( m^2 R^2 + \tfrac{1}{2} \right)\\
&\, + \frac{\beta}{2 R} \sum_{k=1}^{\infty} (-1)^k \left[ \frac{3}{2} \frac{m^2 R^2}{\pi^2 k^2} {\rm K}_2 (2 \pi k m R) + (m^2 R^2 + \tfrac{1}{4}) \frac{m R}{\pi k} {\rm K}_1 (2 \pi k m R) \right]\\
&\, + \frac{\beta}{8 R} m^2 R^2 \bigg[ - \frac{1}{4} + \frac{1}{2} \gamma_{E} + \frac{1}{8} (4 \gamma_E - 3) m^2 R^2 + \frac{1}{2} (1+m^2 R^2) \log \left(\tfrac{m \beta}{\pi}\right) \bigg] \bigg{\}}\\
&\, + d_{{\cal R}} \xi_f (m R, \tfrac{\beta}{R}) \, ,
\label{sumf_apr}
}
where $\xi_f (m R, \frac{\beta}{R}) = {\cal O}\left(\frac{\beta}{R} \right)^3 {\cal O} (m R)^4$ and takes the form
\SP{
&\xi_f \left(m R, \tfrac{\beta}{R}\right)\\
&= \frac{m^2 \beta R}{8} \sum_{n=1}^{\infty} (-1)^n \frac{\Gamma(2n+1) \zeta(2n+1)}{\Gamma(n+3)\Gamma(n+1)} \left( \frac{m \beta}{4 \pi} \right)^{2 n} \left( 2^{2 n + 1} - 1 \right) (2 m^2 R^2 + 2 + n) \, .
}
The contribution to the partition function resulting from the second line in (\ref{sumf_apr}) cancels with the Casimir contribution, and, for regularization scheme $I$, so does part of the third line. The remaining high-temperature partition function, first assuming scheme $I$ regularization of the Casimir energy, is obtained from (\ref{logZf}) using the Casimir energy in (\ref{fercas1}), which gives
\SP{
\log Z_f^{I} \simeq &N_f d_{{\cal R}} \bigg[ \frac{7 \pi^4 R^3}{180 \beta^3} - \frac{\pi^2 R}{12 \beta} (m^2 R^2 + \tfrac{1}{2}) + \frac{\beta}{8 R} m^2 R^2 (1 + m^2 R^2) \left( 1 + \log \left( \frac{\pi R}{2 \beta} \right) \right)\\
&- \xi_f \left(m R, \tfrac{\beta}{R}\right) \bigg] \, .
}
It is now possible to consider whether the Bekenstein bound is satisfied. Calculating $L$ from (\ref{Lbek}) gives
\SP{
L_f^{I} \simeq &N_f d_{{\cal R}} \bigg[ \frac{7 \pi^5 R^4}{30 \beta^4} - \frac{7 \pi^4 R^3}{45 \beta^3} + \frac{1}{12} \left( 1 + 2 m^2 R^2 \right) \left( - \frac{\pi^3 R^2}{\beta^2} + \frac{\pi^2 R}{\beta} \right)\\
&+ \frac{1}{8} m^2 R^2 \left( 1 + m^2 R^2 \right) \left( - \frac{\beta}{R} + 2 \pi \log \left( \frac{2 \beta}{\pi R} \right) \right) + \xi_f' \left(m R, \tfrac{\beta}{R}\right) \bigg] \, ,
\label{Lf1}
}
where $\xi_f' (m R, \frac{\beta}{R}) = {\cal O}\left(\frac{\beta}{R} \right)^2 {\cal O} (m R)^4$ and takes the form
\SP{
\xi_f' \left(m R, \tfrac{\beta}{R}\right) = &\frac{m^2 R}{4} \sum_{n=1}^{\infty} \bigg[ (-1)^n \frac{\Gamma(2n+1) \zeta(2n+1)}{\Gamma(n+3)\Gamma(n+1)} \left( \frac{m \beta}{4 \pi} \right)^{2 n} \left( 2^{2 n + 1} - 1 \right)\\
&\hspace{2cm}\times (2 m^2 R^2 + 2 + n) \left[ (\beta - 2 \pi R) n - \pi R \right] \bigg]\, .
}
To determine the behavior of $L_f$ when $m R \ne 0$ it is necessary to determine if the sum in $\xi_f' (m R, \tfrac{\beta}{R})$ converges. To this end it is helpful to consider the large $n$ limit
\EQ{
\lim_{n \rightarrow \infty} \xi_f' \left(m R, \tfrac{\beta}{R}\right) = \frac{m^2 R}{2 n^{3/2} \sqrt{\pi}} (-1)^n \left( \frac{m \beta}{\pi} \right)^{2 n} \left[ (\beta - 2 \pi R) n - \pi R \right] \, .
}
Setting aside the constants the relevant limits are
\SP{
\frac{(-1)^n}{n^{3/2}} \left( \frac{m \beta}{\pi} \right)^{2 n} \left[ (\beta - 2 \pi R) n - \pi R \right] &\xrightarrow[n \rightarrow \infty]{} \hspace{1cm} 0, \hspace{1cm} {\rm if}~ m \beta \le \pi \, ,\\
&\xrightarrow[n \rightarrow \infty]{} \hspace{5mm} \pm \infty, \hspace{1cm} {\rm if}~ m \beta > \pi \, .
}
Therefore the sum in $\xi_f' \left(m R, \tfrac{\beta}{R}\right)$ converges for $m \beta \le \pi$ and can be truncated as an approximation. However, when $m \beta > \pi$ the sum diverges to $\pm \infty$ depending on whether it is truncated at an even or odd number, and $\xi_f'$ must be regularized.

\begin{figure}[t]
  \hfill
  \begin{minipage}[t]{.49\textwidth}
    \begin{center}
\includegraphics[width=0.99\textwidth]{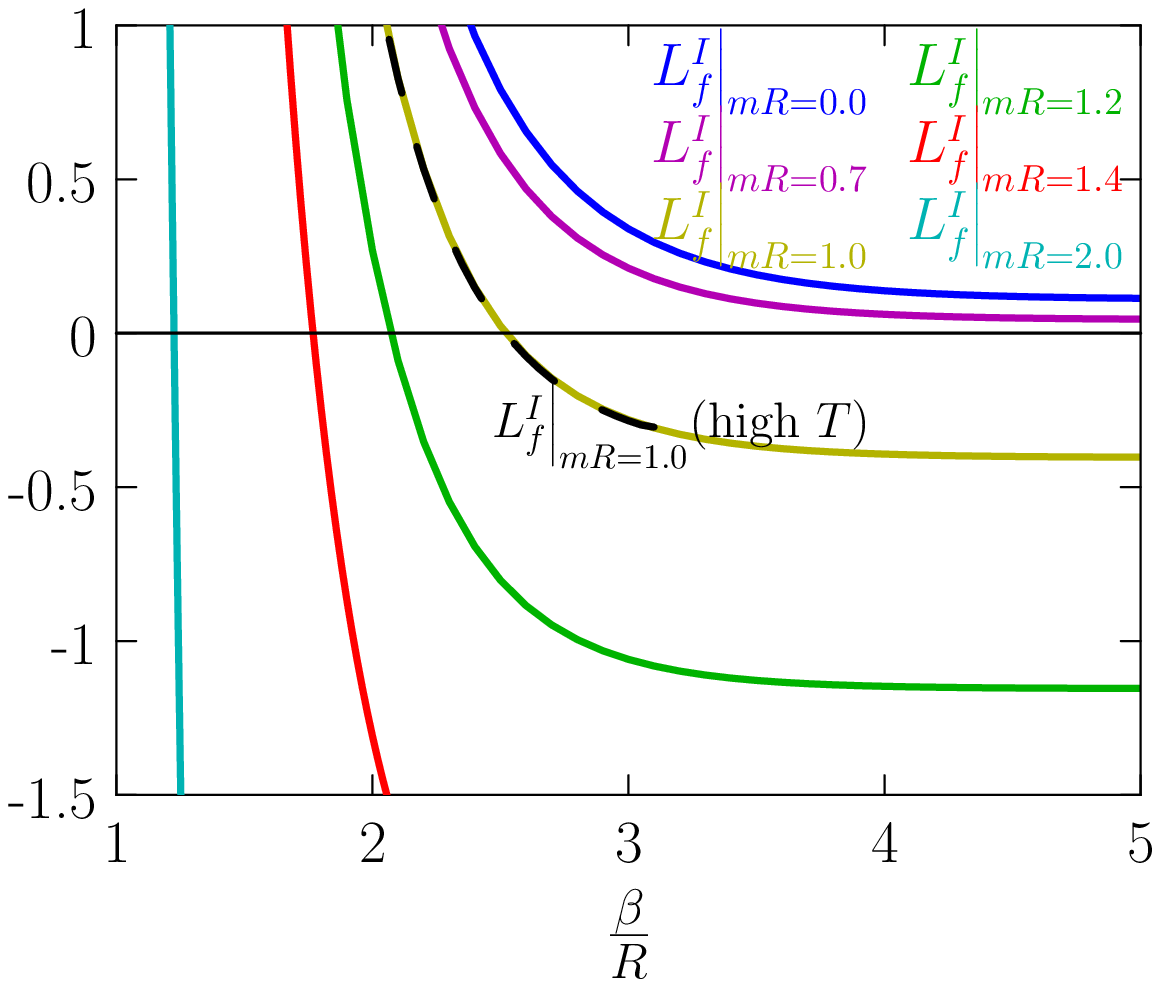}
    \end{center}
  \end{minipage}
  \hfill
  \begin{minipage}[t]{.49\textwidth}
    \begin{center}
\includegraphics[width=0.99\textwidth]{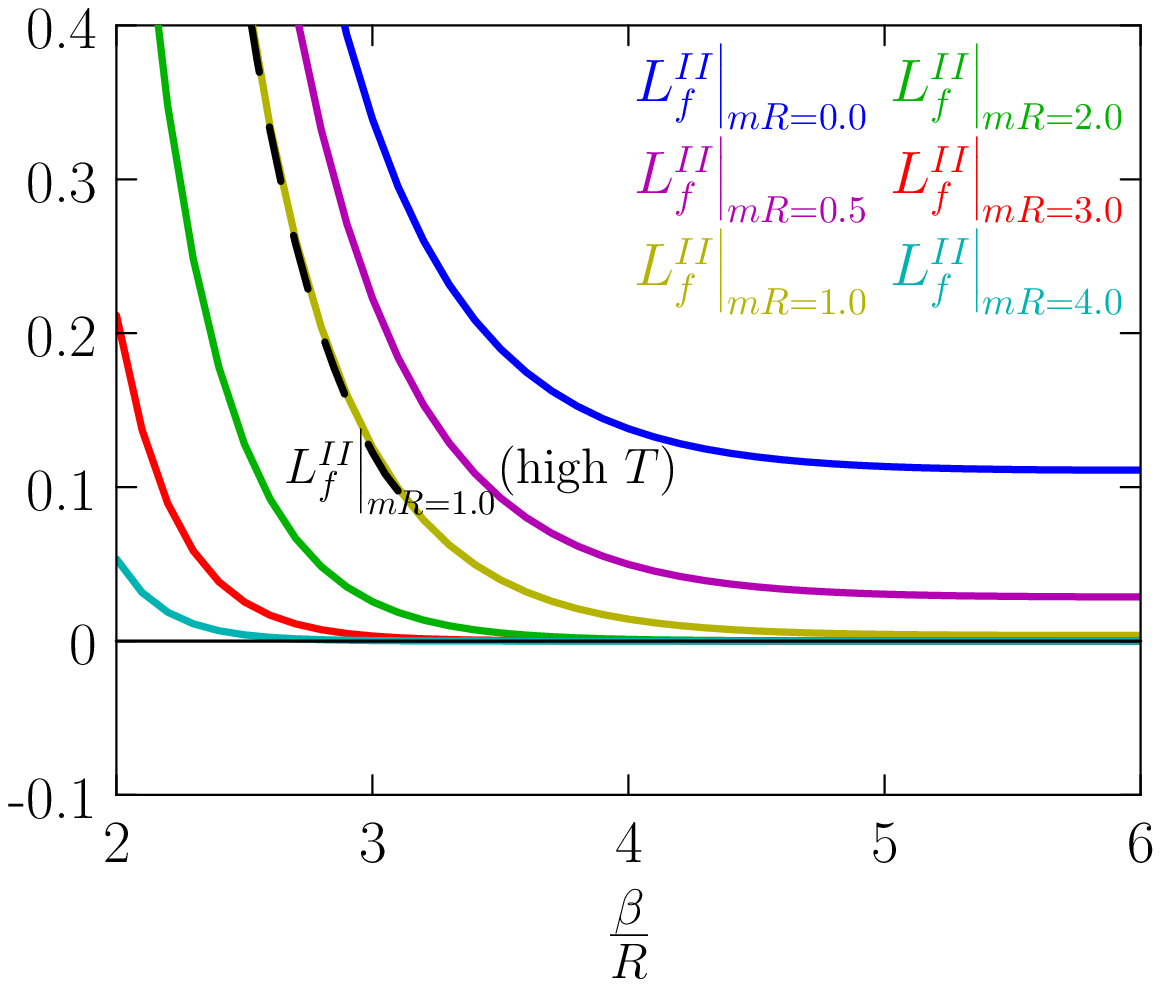}
    \end{center}
  \end{minipage}
  \hfill
\caption{$L$ for fermions with various masses as a function of $\frac{\beta}{R}$ including the low temperature contributions. (Left) $L_f$ calculated including scheme $I$-regularized Casimir energy with $\mu \equiv e^{-\gamma_E}$. (Right) $L_f$ calculated including scheme $II$-regularized Casimir energy with $\mu \equiv \frac{1}{2} m R \exp \left[ \frac{2 + m^2 R^2}{4(1+m^2 R^2)} \right]$. The dotted line gives the high $T$ results for $L_f$ with $m R = 1.0$.} 
\label{plot_Lfm}
\end{figure}

Figure \ref{plot_Lfm} (Left) shows $L_f^I$ in (\ref{Lf1}), labeled $L_f^I ({\rm high}~T)$, as a function of $\frac{\beta}{R}$ for $m R = 1.0$, along with $L_f^I$ computed numerically including the low temperature contributions in (\ref{fer_sum_all}) for various $m R$. Using regularization scheme $I$, with $\mu \equiv e^{- \gamma_E}$, to obtain the Casimir energy, taking $m R \gapprox 0.757$ results in violation of the Bekenstein bound above a certain critical value of $\frac{\beta}{R}$, which decreases with increasing $m R$.

Considering regularization scheme $II$, with $\mu \equiv \frac{1}{2} m R \exp \left[ \frac{2 + m^2 R^2}{4(1+m^2 R^2)} \right]$, the Casimir energy in (\ref{fercas2}) is used with the result (\ref{sumf_apr}) to obtain the partition function via (\ref{logZf}), resulting in
\SP{
\log Z_f^{II} \simeq &N_f d_{{\cal R}} \bigg[ \frac{7 \pi^4 R^3}{180 \beta^3} - \frac{\pi^2 R}{12 \beta} (m^2 R^2 + \tfrac{1}{2}) + \frac{\beta}{8 R} m^2 R^2 \left( \tfrac{1}{2} - \gamma_E + m^2 R^2 \left( \tfrac{3}{4} - \gamma_E \right) \right)\\
&+ \frac{\beta}{8 R} m^2 R^2 \left( 1 + m^2 R^2 \right) \log \left( \frac{\pi}{m \beta} \right) - \xi_f \left(m R, \tfrac{\beta}{R}\right) \bigg] \, .
}
Using this in (\ref{Lbek}) gives
\SP{
L_f^{II} \simeq &N_f d_{{\cal R}} \bigg[ \frac{7 \pi^5 R^4}{30 \beta^4} - \frac{7 \pi^4 R^3}{45 \beta^3} + \frac{1}{12} \left( 1 + 2 m^2 R^2 \right) \left( - \frac{\pi^3 R^2}{\beta^2} + \frac{\pi^2 R}{\beta} \right)\\
&+ \frac{1}{8} \pi m^2 R^2 \left( 1 + \frac{1}{2} m^2 R^2 \right) + \frac{1}{4} \pi \gamma_E m^2 R^2 \left( 1 + m^2 R^2 \right)\\
&+ \frac{1}{8} m^2 R^2 \left( 1 + m^2 R^2 \right) \left( - \frac{\beta}{R} + 2 \pi \log \left( \frac{m \beta}{\pi} \right) \right) + \xi_f' \left(m R, \tfrac{\beta}{R}\right) \bigg] \, .
}
Figure \ref{plot_Lfm} (Right) shows this result, labeled $L_f^{II} ({\rm high}~T)$, for $m R = 1.0$, along with numerical calculations of $L_f^{II}$ including the low temperature contributions from (\ref{fer_sum_all}) for various $m R$. Thus, when the Casimir energy is obtained using regularization scheme $II$, such that $E_{Cas} > 0$ for all $m R$, the Bekenstein bound appears to be satisfied for all $\frac{\beta}{R}$, for any $m R$.

The difference between $L^{II}$ and $L^{I}$ is
\EQ{
L^{II} - L^{I} \simeq \frac{\pi}{4} m^2 R^2 \left[ \frac{1}{2} + \gamma_E + \log \left( \frac{m R}{2} \right) \right] + \frac{\pi}{4} m^4 R^4 \left[ \frac{1}{4} + \gamma_E + \log \left( \frac{m R}{2} \right) \right] \, ,
}
which is positive for $m R \gapprox 0.745$. This reveals that, assuming $L_f^{II} > 0$, it is possible for $L_f^{I}$ to violate the bound for masses above this value, and supports the numerical result that $L_f^I < 0$ for $m R \gapprox 0.757$.
\section{Low temperature partition functions (confined phase)}
\label{lowTconf_sec}

Another interesting question is whether the Bekenstein bound continues to hold if the temperature is decreased and the theory undergoes a phase transition from the deconfined phase to the confined phase. During the transition the eigenvalues of the Polyakov line go from being clumped with $\theta_i = 0$ to being uniformly distributed around the unit circle. For $U(N_c)$ or $SU(N_c)$ theories with odd $N_c$ the Polyakov line angles in the confined phase take the form $\theta = \{ 0, \frac{2 \pi}{N_c}, \frac{4 \pi}{N_c}, ... , \frac{(N_c-1)\pi}{N_c} \}$ (for $N_c$ even the angles are shifted by $\frac{\pi}{N_c}$). Taking the trace of the Polyakov line with the angles evenly distributed around the unit circle results in ${\mathscr P}_n^{({\cal F})} = \sum_{i=1}^{N_c} e^{i n \theta_i} = 0$.

For $U(N_c)$ theories, or $SU(N_c)$ theories with large $N_c$, containing only fundamental and/or adjoint matter, ${\mathscr P}_{n}^{({\cal F})} = 0$ implies that the zero-point contribution to the partition function is all that remains. Since the Casimir energy is always positive for theories with massless matter then the Bekenstein bound is trivially satisfied in the confined phase since $L = -\frac{2 \pi R}{\beta} \log Z = 2 \pi R E_{Cas}$. For the same reason it is also satisfied for massive matter when the Casimir energy is regularized according to scheme $II$. However, for massive matter with the Casimir energy regularized according to scheme $I$ there is a range of $m R$ which results in $E_{Cas} < 0$ and causes the Bekenstein bound to be violated.

The only remaining case where it is not clear if the Bekenstein bound holds is for $SU(N_c)$ theories in the confined phase and with $N_c$ not large. In the next section we consider what happens in pure Yang-Mills theory.
\subsection{$SU(N_c)$ Yang-Mills theory with finite $N_c$}

\begin{figure}[t]
  \hfill
  \begin{minipage}[t]{.49\textwidth}
    \begin{center}
\includegraphics[width=0.99\textwidth]{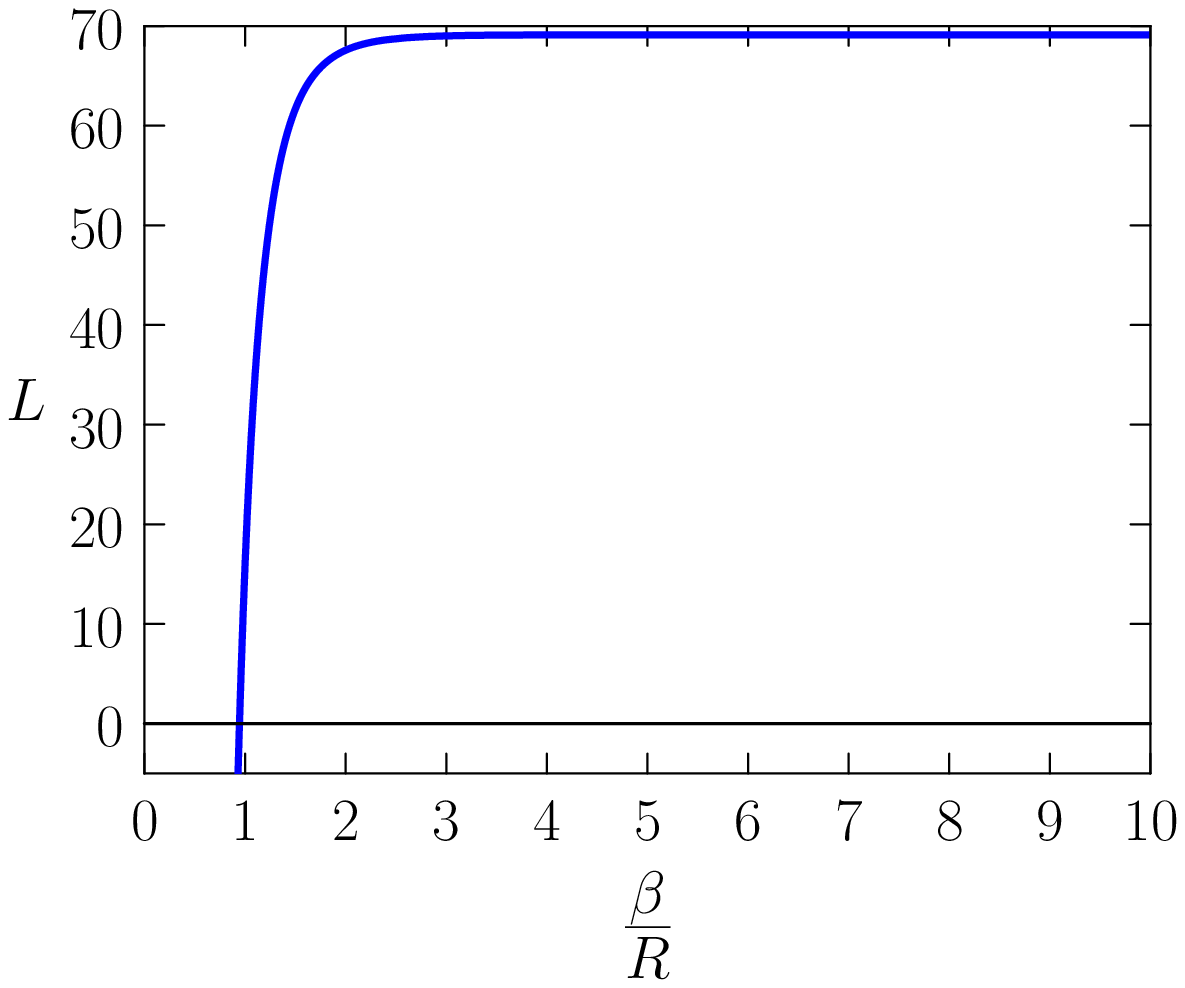}
    \end{center}
  \end{minipage}
  \hfill
  \begin{minipage}[t]{.49\textwidth}
    \begin{center}
\includegraphics[width=0.99\textwidth]{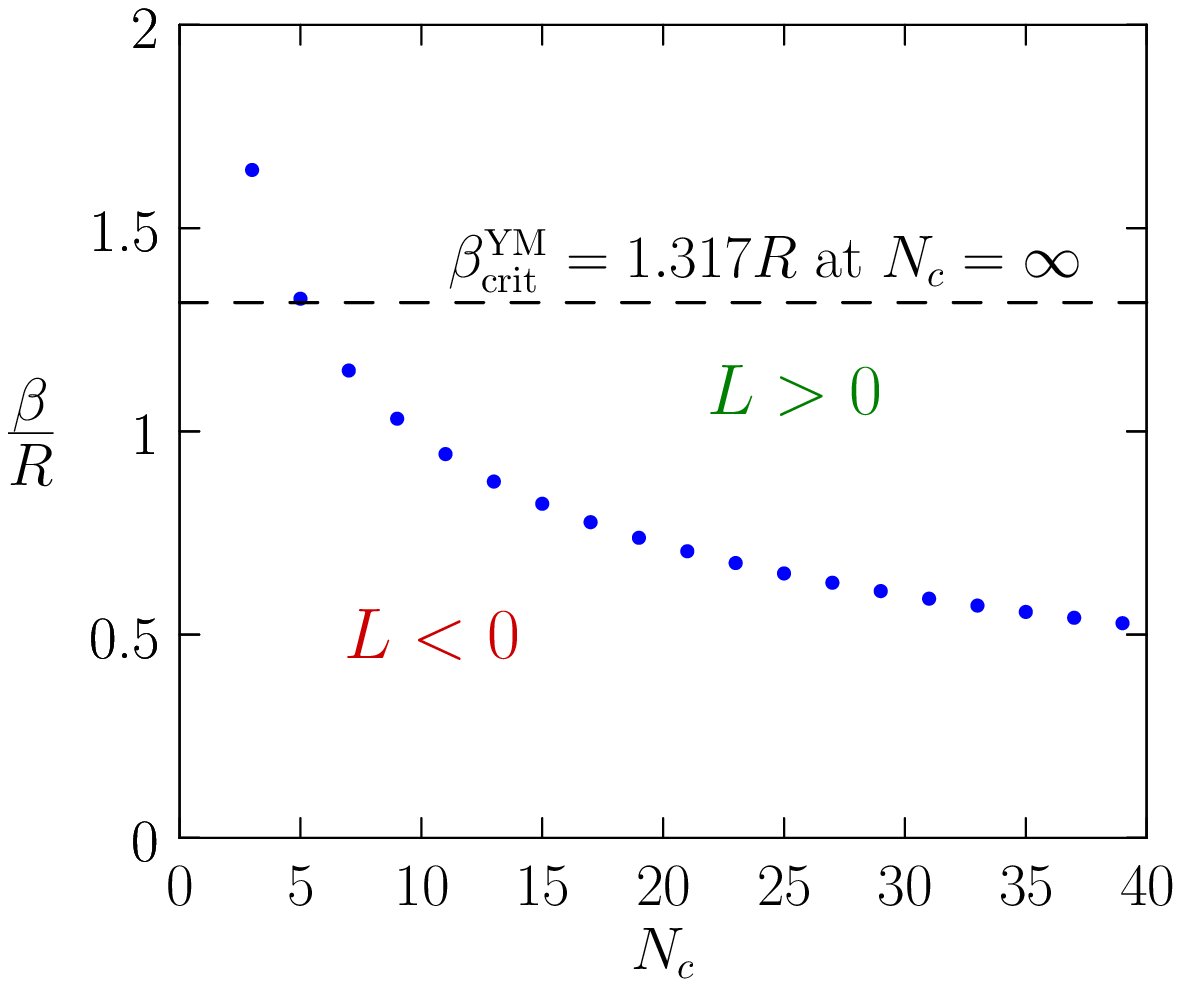}
    \end{center}
  \end{minipage}
  \hfill
\caption{$L$ for theories with massless vectors in the confined phase. (Left) $L$ vs. $\frac{\beta}{R}$ with $N_c = 11$. (Right) Critical $L = 0$ line as a function of $\frac{\beta}{R}$ and odd $N_c$ from $3$ to $39$. The dashed line indicates the temperature of the deconfinement transition for $N_c = \infty$ Yang-Mills theory.} 
\label{plot_Lvec_con}
\end{figure}

For $SU(N_c)$ Yang-Mills theory in the confined phase $\sum_{i,j=1}^{N_c} \cos(n(\theta_i - \theta_j)) = 0$ and it is only necessary to keep the $-1$ contribution from trace over the Polyakov lines in the adjoint representation. The vector sum takes the form
\EQ{
\Xi_{SU(N)}^{YM} = -2 \sum_{n=1}^{\infty} \frac{1}{n} \sum_{l=1}^{\infty} l (l+2) e^{-n \frac{\beta}{R} (l+1)} \, .
\label{ym_sun_sum}
}
In the limit of large $\frac{\beta}{R}$ where the uniform distribution is preferred it is possible to approximate the sum over $n$ by its $n=1$ term,
\SP{
\Xi_{SU(N)}^{YM} &\simeq -2 \sum_{l=1}^{\infty} l (l+2) e^{-n \frac{\beta}{R} (l+1)} \, ,\\
&= \frac{-6 e^{-2\beta/R} + 2 e^{-3\beta/R}}{(1 - e^{-\beta/R})^3} \, .
}
Adding the Casimir contribution $-\beta E_{Cas}^{(v)}$ from (\ref{vec_cas}) the partition function is
\EQ{
\log Z \simeq \frac{-6 e^{-2\beta/R} + 2 e^{-3\beta/R}}{(1 - e^{-\beta/R})^3} - (N_c^2 - 1) \frac{11 \beta}{120 R} \, .
}
Plugging this into (\ref{Lbek}) results in
\SP{
L = &\frac{1}{(1 - e^{-\beta/R})^4}\bigg[ (N_c^2 - 1) \left( - \frac{11 \pi}{60} + \frac{11 \pi e^{-\beta/R}}{15} \right) + \left( -6 + \frac{251 \pi}{10} - \frac{11 N_c^2 \pi}{10} \right) e^{-2\beta/R}\\
&+ \left( (N_c^2 - 1) \frac{11 \pi}{15} + 8 \right) e^{-3\beta/R} + \left( -(N_c^2 - 1) \frac{11 \pi}{60} - 2\right) e^{-4\beta/R} - \frac{12 \beta e^{-2\beta/R}}{R} \bigg] \, .
}
As $\frac{\beta}{R} \rightarrow 0$ it is clear that $L < 0$ and as $\frac{\beta}{R} \rightarrow \infty$ that $L > 0$. At some intermediate $\frac{\beta}{R}$ is the transition point $L = 0$. Figure \ref{plot_Lvec_con} (Left) shows $L$ as a function of $\frac{\beta}{R}$ for $N_c = 11$ and Figure \ref{plot_Lvec_con} (Right) shows the $L = 0$ transition point as a function of $\frac{\beta}{R}$ for odd $N_c$ from $3$ to $39$. For $N_c = 3$, the $L = 0$ transition point occurs at $\frac{\beta}{R} \approx 1.637$, however it should be noted that the saddle point method used to obtain (\ref{logZv1}) is at best a rough approximation when $N_c$ is not large. Keeping more terms in the sum over $n$ in (\ref{ym_sun_sum}) makes a negligible difference. The deconfinement confinement transition occurs at $\frac{\beta}{R} \approx 1.317$ \cite{Aharony:2003sx} for weak-coupling Yang-Mills theory at large $N_c$, however the transition is smeared out when considering finite $N_c$. Therefore it is not clear in the case of $N_c = 3$ if the $L = 0$ point is within the confined phase or not. However, in the $N_c \rightarrow \infty$ limit the $L = 0$ point gets pushed to $\frac{\beta}{R} = 0$, such that $L > 0$ for all $\frac{\beta}{R} > 0$.


\section{Verlinde bound}
\label{Verlinde_sec}

It is straightforward to use the same techniques applied to the Bekenstein bound analysis to test the Verlinde bound \cite{Verlinde:2000wg}, which takes the form
\EQ{
S \le \frac{2 \pi}{3} R E \, ,
}
for conformal theories on $S^1 \times S^3$. For ${\cal N}=4$ SYM theory which has $6$ real scalars, 1 vector, and $4$ Weyl fermions, the relevant inequality is
\EQ{
6 L_{Ver}^{(s)} + L_{Ver}^{(v)} + 4 L_{Ver}^{(f)} \ge 0 \, ,
\label{n4symineq}
}
where $L_{Ver} = \left( \beta - \frac{2 \pi R}{3} \right) \frac{\partial}{\partial \beta} \log Z - \log Z$, and $Z$ is the partition function (including the Casimir contribution) of the massless scalars, vectors, or fermions. Using the high temperature expansions for the partition functions from the previous sections
\SP{
L_{Ver}^{(s)} &\simeq \frac{2 \pi^5 R^4}{45 \beta^4} - \frac{4 \pi^4 R^3}{45 \beta^3} + \frac{1}{4 \pi^2} \zeta(3) \, ,\\
L_{Ver}^{(v)} &\simeq \frac{4 \pi^5 R^4}{45 \beta^4} - \frac{8 \pi^4 R^3}{45 \beta^3} - \frac{2 \pi^3 R^2}{9 \beta^2} + \frac{2 \pi R}{3 \beta} (\pi + 1) + \frac{1}{2 \pi^2} \zeta(3) - 1 + \log \left(\frac{\beta}{2 \pi R}\right) \, ,\\
L_{Ver}^{(f)} &\simeq \frac{7 \pi^5 R^4}{90 \beta^4} - \frac{7 \pi^4 R^3}{45 \beta^3} - \frac{\pi^3 R^2}{36 \beta^2} + \frac{\pi^2 R}{12 \beta} \, .
\label{Lveraprs}
}
Thus the inequality in (\ref{n4symineq}) takes the form
\EQ{
\frac{2 \pi^5 R^4}{3 \beta^4} - \frac{4 \pi^4 R^3}{3 \beta^3} - \frac{\pi^3 R^2}{3 \beta^2} + \frac{\pi R}{\beta} (\tfrac{2}{3} + \pi) + \frac{2}{\pi^2} \zeta(3) - 1 + \log \left(\frac{\beta}{2 \pi R}\right) \ge 0 \, ,
}
which means that $L > 0$ for $\frac{\beta}{R} \lapprox 1.648$. Since this value is in the regime where the high temperature approximation is very good calculating $L_{Ver}^{(x)}$ in (\ref{Lveraprs}) while including the low temperature corrections makes a negligible difference. We have checked this numerically. Note that the deconfinement-confinement transition at zero 't Hooft coupling occurs at $\frac{\beta}{R} \approx 2.634$ \cite{Aharony:2003sx}, and the corresponding strong coupling Hawking-Page transition occurs at $\frac{\beta}{R} = \frac{2 \pi}{3} \approx 2.094$ \cite{Hawking:1982dh,Witten:1998zw}. Either way, it appears that as $\frac{\beta}{R}$ is increased, the weakly coupled theory violates the bound before the transition takes place. It is however important to recall the Vandermonde contribution $L_{Vdm} \simeq d_{{\cal A}} ( \log \Lambda + \gamma_E )$, which we dropped at the end of Section \ref{logZreview}. It is possible that this contribution is sufficient to prevent violation of the Verlinde bound. The $\theta_i$-dependence, which is relevant near a phase transition, would need to be properly accounted for to be sure.

\section{Conclusions}
\label{conclusions_sec}

Using the approach in \cite{Cardy:1991kr,Kutasov:2000td} we have calculated the high temperature partition functions for gauge theories with massless vectors, scalars, and/or fermions, and for gauge theories with massive scalars and/or fermions, at zero 't Hooft coupling on $S^1 \times S^3$, in the deconfined phase. From these results, and from numerical results including the low temperature contributions, the Bekenstein entropy bound was determined to hold at all $\frac{\beta}{R}$ for theories with massless deconfined matter. This was also determined to be true for theories with massive scalars or fermions regularized using scheme $II$ to obtain the Casimir energy, with regularization parameter $\mu$ fixed to satisfy the constraint $E_{Cas} \rightarrow 0$ as $m R \rightarrow \infty$. When the Casimir energy was instead regularized using scheme $I$, which fixes $\mu$ to match the zeta function result with the cutoff method result, then for both scalars and fermions we found a range of $m R \ne 0$ for which the Bekenstein bound was violated for a range of $\frac{\beta}{R}$.

Since it is clear that the Casimir energy must be positive in order for the Bekenstein bound to hold all the way into the low temperature regime then only our scheme $II$ results could be expected to hold at all temperatures, and the fact that they appear to hold regardless of $m R$ is consistent with Bekenstein's conjecture. What is yet unclear is which regularization procedure gives the correct, scheme-independent Casimir energy for theories with massive scalar or fermionic matter. If it is scheme $I$ that gives the correct results then for theories with scalars or fermions of sufficiently low $m R$ Bekenstein's bound still appears to hold at all temperatures.

We also considered matter in the confined phase to determine if the Bekenstein bound still holds when the theory undergoes a phase transition. In the confined phase the $\Tr_{{\cal R}} e^{i n \theta}$-dependent contributions to the partition function vanish in the case of $U(N_c)$ theories or $SU(N_c)$ theories at large $N_c$ when the matter is in the adjoint and fundamental representations. The only case where it is unclear if the Bekenstein bound holds in the confined phase is for $SU(N_c)$ theories with $N_c$ not large. It might be possible to check this numerically by performing the integrals over the $\theta_i$ directly rather than using the saddle point approximation to get from the action to the partition function.

With regard to the Verlinde bound we found, in agreement with \cite{Kutasov:2000td}, that it is still violated for sufficiently large $\frac{\beta}{R}$ before the deconfinement-confinement transition takes place, when considering ${\cal N} = 4$ SYM as a free theory. However, for the $SU(N_c)$ theory at zero 't Hooft coupling, the Vandermonde contribution could be sufficient to prevent violation of the bound, but the $\theta_i$-dependence of the partition function near the transition temperature would need to be included to be sure.
\section{Acknowledgements}

We would like to thank Jan Rosseel for many useful conversations, for looking at the manuscript, and for checking some of the results. We would also like to thank Jose Barbon for helpful discussions regarding the physical relevance of the Bekenstein bound. The research of JCM is supported by the Stichting voor Fundamenteel Onderzoek der Materie (FOM).

\clearpage

\end{document}